\documentclass[sigplan, 10pt, nonacm]{acmart}
\thanks{$^*$Equal contribution.}

\usepackage{tcolorbox}
\usepackage{xspace}
\usepackage{outlines}

\usepackage{caption}
\newcommand{\parab}[1]{\vspace{0.03in}\noindent{\bf #1}}

\newcommand{\subheading}[1]{\vspace{2pt}\textit{#1:}}

\usepackage{enumitem}

\definecolor{lightgray}{gray}{0.92}
\newcommand\greybox[1]{%
\vspace{6pt}%
      \par{\centering\colorbox{lightgray}{%
              \begin{minipage}{3.3in}#1\end{minipage}%
                        }%
                                    \vskip 2pt%
                                    \vspace{1pt}%
                                                }}

\newcommand{\fix}{\textcolor{black}}

\newcommand{\router}{{\sc XWind}\xspace}
\newcommand{\greenf}{{Greeninferencing}\xspace}

\newcommand{\slc}{{\sc XW-Slc}\xspace}

\newcommand{\azure}{{\sc Azure}\xspace}

\makeatletter
\renewcommand{\@thanks}{\footnote{\@gobbleto@sep\thanks@a}}
\def\thanks@a{\let\@thanks\relax\footnotetext}
\makeatother

\usepackage{graphicx}
\usepackage{subcaption}
\usepackage{float}

\usepackage{algorithm}
\usepackage{algorithm}
\usepackage{algpseudocode}
\usepackage{enumitem}

\algnewcommand{\LineComment}[1]{\Statex \hspace{\algorithmicindent} \(\triangleright\) #1}
\algrenewcommand\algorithmicrequire{\textbf{Input:}}
\algdef{SE}[SUBALG]{Indent}{EndIndent}{}{\algorithmicend\ }

\algblockdefx[CLASS]{Class}{EndClass}[1]{\textbf{class} \textsc{#1}}{\textbf{end class}}

\usepackage[utf8]{inputenc}
\usepackage{amsmath}
\usepackage{placeins}

\newenvironment{myitemize}
{\begin{itemize}[leftmargin=1.75em]
  \setlength{\itemsep}{2pt}
  \setlength{\parskip}{0em}  
  \setlength{\parsep}{0pt}
  \vspace{-3pt}}
{\vspace{-3pt}\end{itemize}}
\newenvironment{myenum}
{\begin{enumerate}[leftmargin=1.75em]
  \setlength{\itemsep}{2pt}
  \setlength{\parskip}{0em}
  \setlength{\parsep}{0pt}
  \vspace{-6pt}}
{\vspace{-6pt}\end{enumerate}}

\definecolor{lightgray}{gray}{0.92}
\title{\router{}: A Cross-site Router for Large Language Model Inference Serving at Renewable Energy Farms}

\begin{document}
\sloppy
\raggedbottom
\author{%
  \fontsize{10pt}{12pt}\selectfont{}Tella Rajashekhar Reddy*, 
  \fontsize{10pt}{12pt}\selectfont{}Atharva Deshmukh*,
  \fontsize{10pt}{12pt}\selectfont{}Liangcheng Yu,
  \fontsize{10pt}{12pt}\selectfont{}Chaojie Zhang,
  \fontsize{10pt}{12pt}\selectfont{}Mike Shepperd,
  \fontsize{10pt}{12pt}\selectfont{}Rohan Gandhi,
  \\
  \fontsize{10pt}{12pt}\selectfont{}Anjaly Parayil,
  \fontsize{10pt}{12pt}\selectfont{}Srinivasan Iyengar,
  \fontsize{10pt}{12pt}\selectfont{}Ajay Manchepalli,
  \fontsize{10pt}{12pt}\selectfont{}Debopam Bhattacherjee
}
\affiliation{%
\vspace{0.05in}
  \fontsize{10pt}{12pt}\selectfont{}Microsoft
  \country{}
}

\begin{abstract}
AI power demand is growing at an unprecedented rate while power grids are often ailing and struggle to keep up. Grid expansion comes with high capital expenditure and long-distance transmission losses, yet there is \textit{abundant} renewable energy at the source, just not matched to demand. 

This paper proposes a complementary AI infrastructure deployment model, \textit{AI \greenf{}}, that brings modular AI compute to renewable energy sources, focusing on wind, allowing AI footprint expansion, generating local behind-the-meter demand for renewable sites, and helping ease the growing strain on power utilities. 
Our feasibility analysis shows that $890+$~GW of wind capacity lies within $50$~ms network round trip time of \azure{} data centers, and that site-wise right-sizing combined with spatial complementarity of wind energy keeps aggregate fleet utilization on par with traditional deployments.

To serve inference requests under variable wind power, we build \router{}, a lightweight, reactive, and workload-agnostic AI inference router that uses only real-time signals: inference latency, KV-cache utilization, and queue depth, to dynamically configure sites and distribute requests. 
Evaluated on a real $64$-GPU A$100$ testbed emulating three wind-powered sites with \azure{} production traces, \router{} reduces P$99$ end-to-end latency by up to $52\%$ over the strongest contender (also our idea) and by up to $98\%$ over baselines such as power-capping and GPU idling, with consistent gains across workload types, load levels, and GPU generations.

\end{abstract}

\maketitle
\thispagestyle{plain}
\pagestyle{plain}


\section{Introduction}

AI adoption is accelerating across the industry, governments, and individuals~\cite{stanford_ai_index_2025, github_copilot, bick2024rapid}, and so is the energy bill. The IEA estimates~\cite{iea_energy_ai_2025} that global data center electricity consumption reached $415$~TWh in $2024$, roughly $1.5\%$ of global demand, and projects it to more than double to $945$~TWh by $2030$, on par with Japan's electricity use. A key driver is the increasing power density of AI hardware~\cite{power_density_1, power_density_2, nvidia_superpod}, with rack scale demands shooting up well above $100$~kW, thus significantly inflating data center scale demands. AI inferencing, which accounts for $90\%$ of AI compute today~\cite{semianalysis_report, chernicoff2024aiworkloads}, is the dominant and fastest-growing segment of this demand.

Seeing this surge, hyperscalers have announced partnerships~\cite{ms_energy_deals, google_20B_deal, ms_three_mile, reuters2025_amazon_ai} with energy providers to secure power. Unfortunately, there isn't a silver bullet that addresses the growing demand and ailing delivery. First, expanding grid infrastructure: new transmission lines, distribution systems, or energy storage is capital-intensive~\cite{ fares2017trends, TnD_cost, eei_grid_investment_2025}, often faces regulatory and logistic delays~\cite{ berkeley_lab_queued_up_2025}, and is especially difficult when renewable sources are located far from consumption hubs~\cite{wind_far, china_curtailment}. A recent Berkeley Lab report~\cite{berkeley_lab_queued_up_2025} highlights that by $2024$, pending grid approvals for new power generation exceeded twice the installed US capacity, with a median wait time of $4.5$--$5$ years. Second, even approved projects often face curtailment due to grid congestion, leaving clean, already generated power underutilized~\cite{eia_curtailment, amperon_curtailment_2024, uk_curtailment}. Third, long-distance transmission and distribution (T\&D) losses significantly inflate the cost of electricity~\cite{fares2017trends, TnD_cost}. For example, the EIA reports~\cite{eia_electric_power_monthly_2026} US industrial rates of $9.3$~\textcent/kWh, while wind farms sell at $2.3$-$4.5$~\textcent/kWh~\cite{nrel_wind_cost_2024, lazard_lcoe_2025} at the source. Finally, much of the grid infrastructure is aging~\cite{mckinsey_eu_ai_power, berkeley_lab_queued_up_2025} and operators may be resorting to short-term measures~\cite{dominion_energy, evergy} that compromise long-term sustainability.

Several strategies have emerged to address these concerns. On-site fossil generation, such as gas turbines and fuel cells~\cite{crusoe_abilene_2026, ge_vernova_crusoe_2025, bloom_energy_wyoming_2026}, offers speed but at the cost of carbon emissions. Nuclear energy promises clean baseload, with hyperscalers signing multi-gigawatt deals~\cite{constellation_tmi_2025, doe_tmi_loan_2025, google_kairos_2024}, but most deployments face regulatory hurdles and construction timelines stretching beyond $2030$~\cite{introl_nuclear_ai_2025} while the demand is \textit{real} and \textit{now}. Space-based computing~\cite{starcloud_orbital_2025, techrepublic_space_dc_2025} leverages perpetual solar energy above clouds, but remains an early-stage, costly demonstration with significant deployment challenges.

This work is motivated by a more immediate lever: consuming renewable energy where it is generated \textit{today}. Renewable energy farms produce power that could benefit from local on-site demand to
cover grid uncertainties (interconnection queues~\cite{berkeley_lab_queued_up_2025} and curtailment~\cite{amperon_curtailment_2024, eia_curtailment_california_2024, modo_energy_ercot_2024}). Co-located
compute startups~\cite{windcores, soluna, westfalenwind} have begun deploying at such sites for crypto-mining and content streaming, but LLM inferencing, the dominant and fastest-growing AI workload,
presents a far larger and yet unexplored opportunity.

\begin{figure}[t]
    \centering
    \includegraphics[width=0.95\columnwidth]{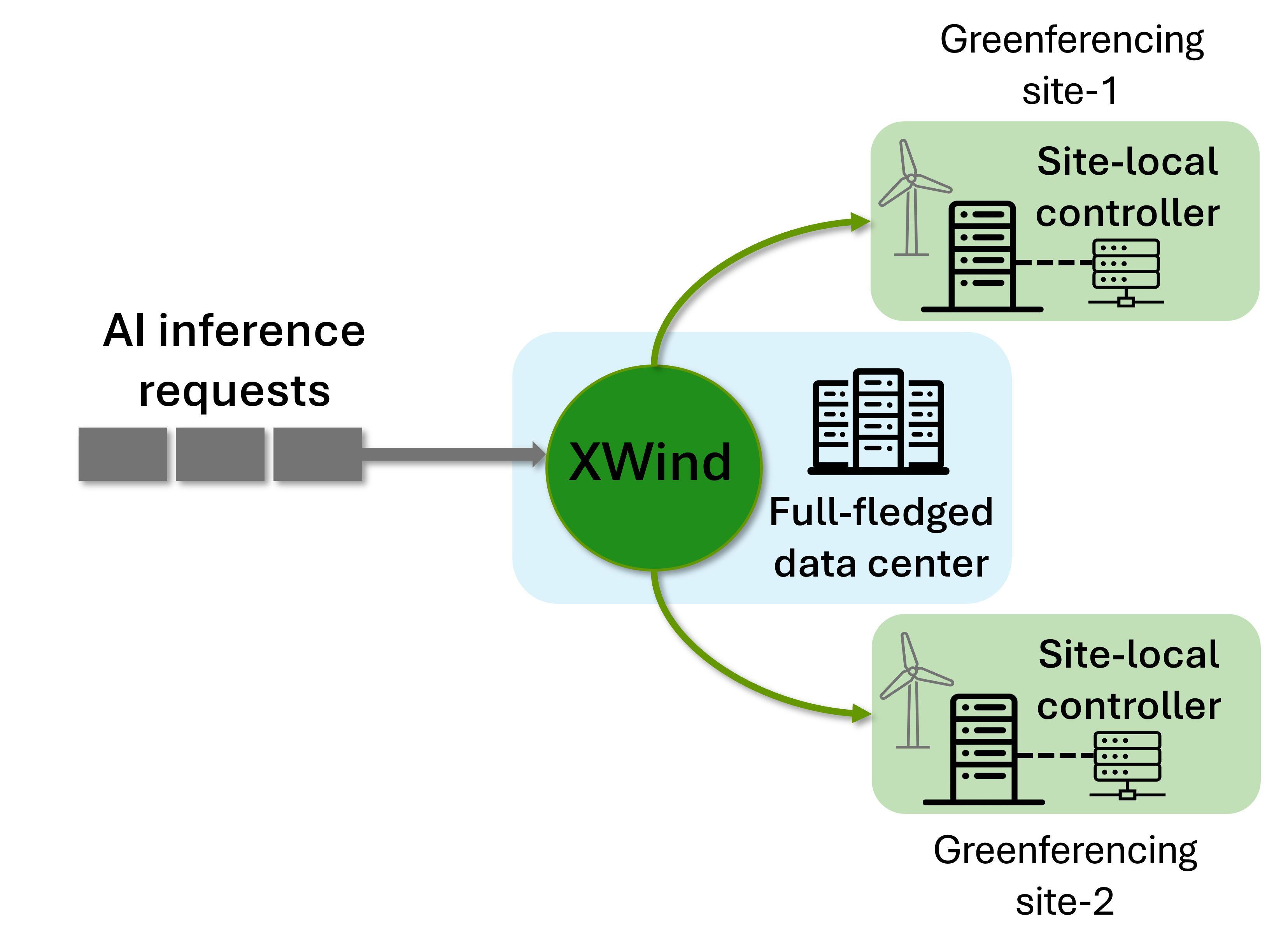}
    \vspace{-0.2in}
    \caption{AI \greenf{} with \router{}.} 
    \label{fig:xwind-architecture}
\end{figure}

\parab{AI \greenf{}.} We propose AI \greenf{} that co-locates modular AI compute deployments at existing or upcoming (also, otherwise queued) renewable energy sites (to start with, wind farms), as seen in Fig.~\ref{fig:xwind-architecture} (green boxes on the right), and helps run a significant share of AI inferencing workload sustainably at a lower cost of energy (no T\&D loss or CAPEX). 
This deployment strategy aligns with the growing need for `community-first' AI infrastructure~\cite{brad_smith} that benefits host communities. AI Greenferencing is broadly a win-for-all: ($1$)~users gain access to sustainable AI services; ($2$)~AI providers unlock additional compute capacity, user reach, and revenue; ($3$)~wind farms can monetize output locally; and ($4$)~power grids benefit from reduced load and a breather for expansion. Note that \greenf{} is designed not to obviate but to co-exist with traditional data centers as a complementary deployment model. Although probably intuitive in hindsight, realizing AI \greenf{} requires navigating several practical concerns that we discuss below.


\parab{Enough wind power to make a dent?} We find that there is already a significant wind capacity (only $100+$~MW large deployments considered, operating and under-construction) of $890+$~GW globally, as per the Global Energy Monitor data~\cite{wind_data_global_energy_monitor} within $50$~ms fiber (circuitous routes, slower speed-of-light than in air) round-trip time (RTT) of \azure{} DCs. $73\%$ of this capacity is within $20$~ms RTT. While our deployment right-sizing strategy is to deploy compute with peak demand at a low percentile (not the peak) of a site's generation, more than $10$~million NVIDIA H$100$ equivalents could be deployed \textit{today} in wind farms with \greenf{}.

\parab{Power variations a problem?} We find that wind power, although variable, is highly predictable on $15$-min timescales (autocorrelation $>0.99$~\cite{elia, emhires}), thus assisting the scheduler in making informed decisions while routing online. Furthermore, geographically dispersed wind sites exhibit spatial complementarity~\cite{virtual_battery, skybox} in generation. Cross-country combinations reduce the coefficient of variation by up to $36\%$~\cite{emhires} allowing to route around transient power drops. 

\parab{\router{} for AI Greenferencing.} When co-locating GPU farms with wind sites, a key challenge is to serve AI inferencing requests under variable power and workload arrival. Existing energy-efficient schedulers~\cite{dynamollm} assume stable grid power and drop up to $50$\% of requests under power dips in our cross-site experiments, even where none of the drops is necessary with intelligent re-routing. 
To realize the \greenf{} vision, we design \router{} (Fig.~\ref{fig:xwind-architecture}), a lightweight, reactive, and profiling-free power-variability-aware cross-site AI inference router. \router{} works in tandem with site-local controllers (\slc{}) that leverage online telemetry (KV-cache utilization, queue depth, and inferencing latency) to reconfigure sites during rare windows of power crunch. These controllers share useful reconfiguration and telemetry signals with \router{} that could then intelligently re-route around constrained sites. 

\parab{}This paper makes the following contributions:
\vspace{0.1in}
\begin{myitemize}
    \item \textbf{We propose AI \greenf{}}, a regionally geo-distributed deployment model that co-locates AI compute at wind farms behind-the-meter. Our opportunity analysis shows this is not a blip: $10+$~million H$100$ GPU equivalents could be deployed today within a few tens of milliseconds of \azure{} data centers (\S\ref{sec:feasibility}). We are also working closely with a large renewable energy company that sees significant value in this strategy. 
    \item \textbf{We design and build \router{} cross-site router} for \greenf{}. \router{} (\S\ref{sec:xwind}) is lightweight, power-variability-aware, and uses only real-time telemetry signals, while efficiently routing AI inference requests acorss multiple variable-power sites.
    \item \textbf{We evaluate \router{} on a $64$-GPU A$100$ testbed} emulating three wind-powered \greenf{} sites, the first hardware demonstration of inference serving under variable renewable power. \router{} reduces P$99$ end-to-end (E$2$E) latency by $22$--$52\%$ over the strongest contender (also our idea) and by up to $98\%$ over baselines (\S\ref{sec:evaluation}). 
\end{myitemize}


\section{Feasibility Analyses}
\label{sec:feasibility}
AI compute is increasingly power-dense~\cite{power_density_1, semianalysis_report, power_density_3}, with the US data center power demand growing at $10$--$15\%$ CAGR\cite{bcg_report}, approaching a significant fraction of residential consumption. This risks overwhelming aging grid infrastructure and forcing short-term decisions\cite{evergy, dominion_energy} that undermine long-term sustainability. Wind farms offer a massive, yet underutilized alternative. As of February $2026$, the global wind capacity pipeline stands at $3$~TW~\cite{wind_data_global_energy_monitor}. However, much of this capacity remains stranded in interconnection queues\cite{berkeley_lab_queued_up_2025}, and even operating farms face curtailment due to grid congestion~\cite{amperon_curtailment_2024}.

This section quantifies the \greenf{} opportunity that helps AI consume this green energy at its source: \S\ref{subsec:wind-capacity} assesses reachable capacity and economics; \S\ref{subsec:taming-uncertainty} addresses power and workload fluctuations; and \S\ref{subsec:lost-cycles} shows how to minimize lost compute cycles.

\subsection{Wind Capacity and Economic Viability}
 \label{subsec:wind-capacity}
\begin{figure}[t]
    \centering
    \includegraphics[width=0.8\columnwidth]{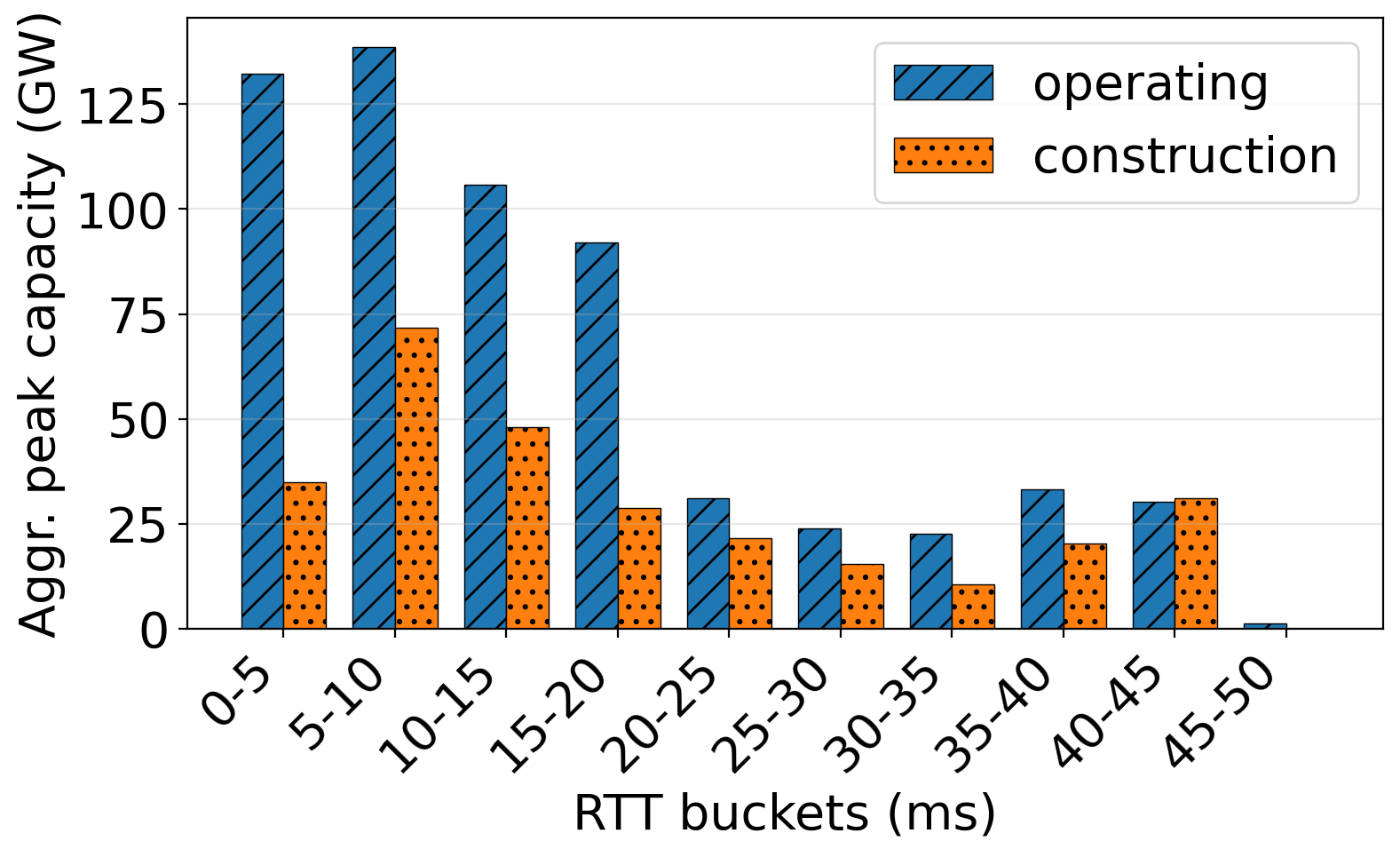}
    \caption{Massive wind capacity (for large $100+$~MW farms) lies within $20$~ms fiber RTT of \azure{} data centers.}
    \label{fig:rtt-capacity}
\end{figure}

The search for wind farms within a reasonable network distance from the data centers is crucial. We need to slightly digress here and first understand the AI inference latency components~\cite{dynamollm} with SLOs (service level objectives): Time To First Token (TTFT, few $100$~ms to seconds) and Time Between Tokens (TBT, lower). E$2$E latency encompasses request queue time, TTFT, and TBT. 
Network latency primarily affects TTFT, not TBT (tokens are streamed). Using the Global Energy Monitor dataset~\cite{wind_data_global_energy_monitor}. we find that $890$+~GW of operating and under-construction wind capacity ($100+$~MW farms only) lies within $50$~ms fiber RTT of \azure{} DCs, with $73$\% within $20$~ms (Fig.~\ref{fig:rtt-capacity}) thus not significantly affecting TTFT (and hence the E$2$E latency).
Similar analysis with 
another hyperscaler's data centers reveals comparable proximity. Being able to tap into even $2\%$ of this generation (often curtailed/queued) unlocks additional capacity larger than today's largest data centers.

\parab{Makes economic sense?} CAPEX is comparable: infrastructure costs have a similar breakeven as modular data centers~\cite{microsoft_mdc} that can be mass-produced at
scale~\cite{mdc_cost_1, mdc_cost_2, mdc_adv}, and GPU costs are identical regardless of location. On the OPEX side, higher maintenance costs from the distributed setup~\cite{dc_maintenance} are easily
offset by $2$-$4\times$ lower power costs at source: the EIA~\cite{eia_electric_power_monthly_2026} reports US industrial rates of $9.3$\textcent/kWh, whereas wind PPAs are $2.3$-$4.5$
\textcent/kWh~\cite{nrel_wind_cost_2024, lazard_lcoe_2025}. Whether utilization at these sites can match traditional deployments is addressed in \S\ref{subsec:lost-cycles}.


\subsection{Operational Feasibility}
\label{subsec:taming-uncertainty}

AI Greenferencing must tame uncertainties at both ends: wind power generation and inference workload arrival. 

\parab{Wind is predictable.} Wind power, albeit variable, is predictable: a characteristic the AI \greenf{} software can leverage when routing workload across wind RE sites. Across $4$ combinations of wind regions (Wallonia,
Flanders) and power grids (Elia, Dso) at $15$min granularity (Jan--Jul'$24$, ELIA~\cite{elia}), the mean autocorrelation at a lag of $1$ is $0.991$. Across $235$ wind farms in the EMHIRES dataset~\cite{emhires}, the mean (median) autocorrelation at a lag of $1$ is $0.99$ over $1$~year ($2018$--$19$) of hourly generation data. These scores confirm strong predictability at different temporal granularities with time series or ML-based models that can consume additional features such as historical data, seasonality, local weather, and turbine specifications. 
The industry uses standard predictors like TFT~\cite{google_tft} (Google) and DeepMC~\cite{kumar2021micro} (Microsoft) for wind power prediction with very high accuracy. The broad availability of such predictors~\footnote{Our in-house framework (orthogonal work) yields a $20\%$ relative improvement in prediction error over state-of-the-art predictors.} helps us treat wind power generation as an oracle (variable yet predictable with high accuracy) in \greenf{} systems design.

\parab{Spatial complementarity smooths variability.} Wind generation across geographically dispersed sites exhibits complementarity, when one site has low wind, others still tend to have high wind intensity. In the EMHIRES dataset, combinations of $4$ cross-country sites (e.g., Iceland, Norway, Switzerland, UK) reduce the coefficient of variation (CoV) of aggregate generation by 36\% compared to a single site~\cite{emhires}. 

\begin{figure}[t]
    \centering
    \includegraphics[width=\columnwidth]{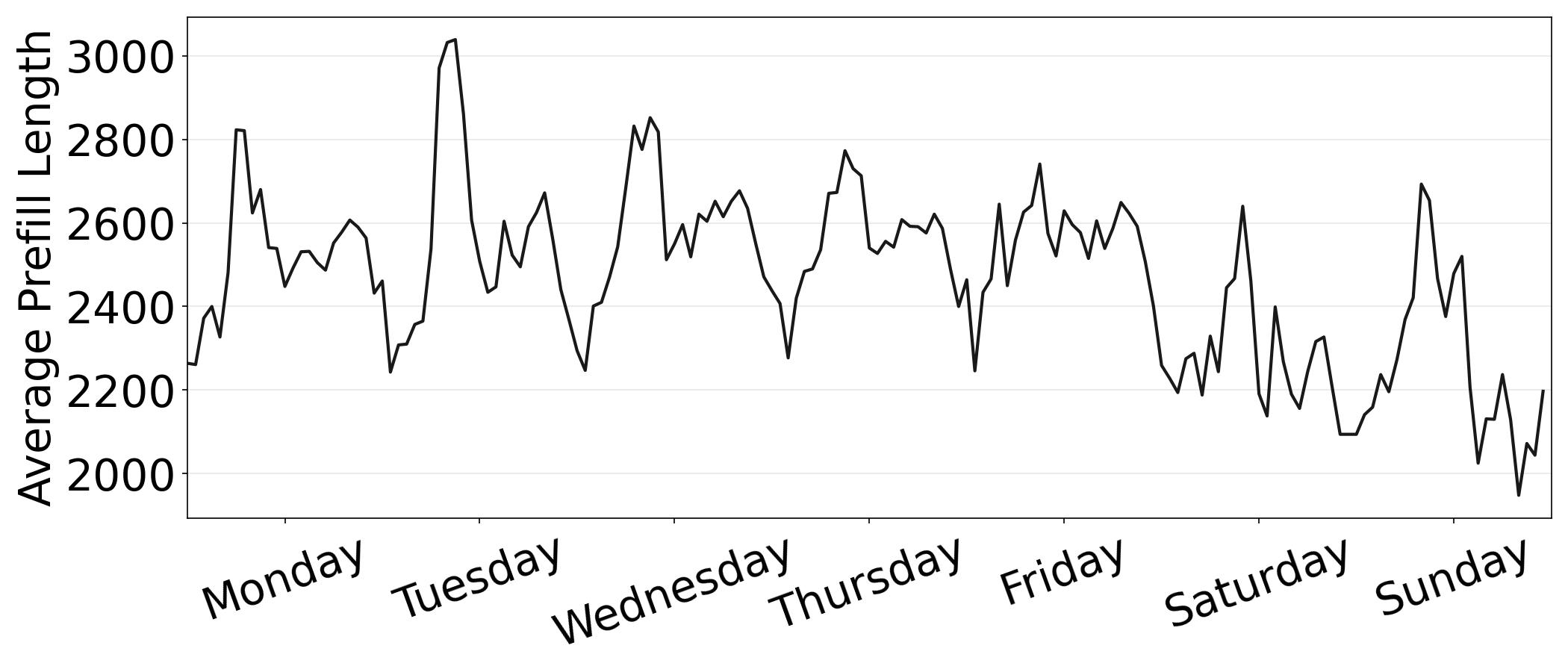}
    \caption{\azure{} coding trace: average prefill length varies significantly over a week.}
    \label{fig:varying-prefills}
    \vspace{-0.2in}
\end{figure}

\parab{Workload is harder to characterize than power.}
Wind power varies continuously but predictably.
In contrast, AI inference workloads exhibit high variability in prefill lengths, decode lengths, and their ratios across workload types, shifting over time (Fig.\ref{fig:varying-prefills}), with arrival patterns disrupted by flash crowds~\cite{altman2025gpus, reuters2025_ghibli_effect}. Proactively predicting workload characteristics, as optimization-based routers require, demands offline profiling per workload type and accurate output-length predictors, both hard to maintain in production. A reactive approach is feasible: memory pressure, latency, and queue buildup are directly observable through runtime telemetry regardless of workload composition. This calls for a design that is \textit{proactive on power} and \textit{reactive on workload}.


\subsection{Compute Feasibility}
\label{subsec:lost-cycles}


Another aspect of deploying GPU compute at these variable power wind sites is the potential loss of some compute cycles. 
Multiple guardrails mitigate this risk: \textit{right-sizing} deploys compute at a conservative percentile (e.g., $20$th) of each site's peak generation; \textit{batteries} bridge transient dips before the router acts; \textit{additional modalities} like solar complement wind; sites can \textit{opportunistically tap the grid} during sustained shortfalls; and \textit{cross-site complementarity} helps redistribute inference load to sites with available power, exploiting the spatial diversity discussed in \S\ref{subsec:taming-uncertainty}.

\begin{figure}[t]
     \centering
     \includegraphics[width=0.9\columnwidth]{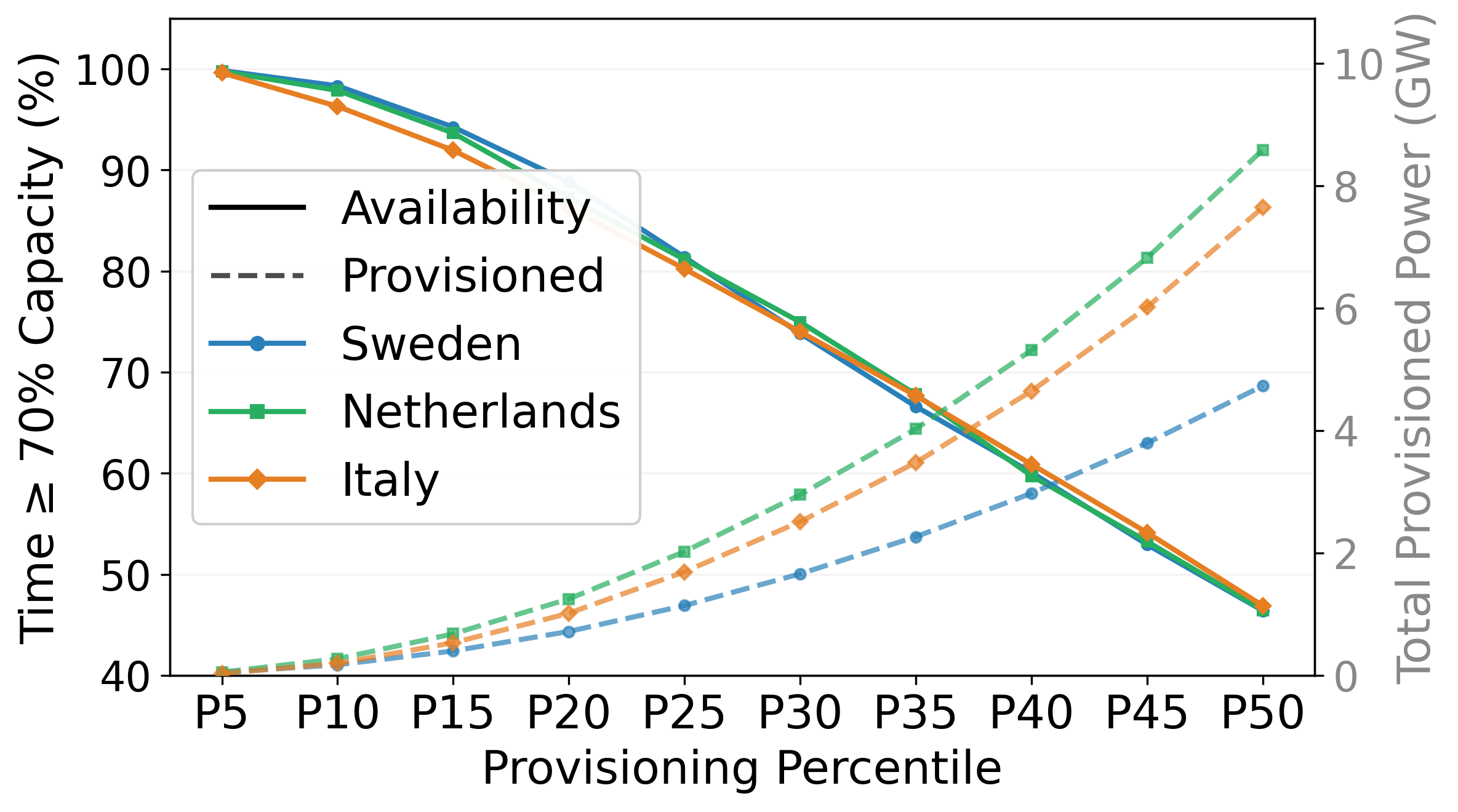}
     \caption{Availability vs.\ provisioning tradeoff for $3$ \azure{} DCs at $20$~ms RTT.} 
     \label{fig:avail-vs-prov}
 \end{figure}

\parab{How often does aggregate power fall short?} To quantify this, we use the EMHIRES dataset~\cite{emhires} with $1$ year of hourly wind generation across European NUTS$2$ regions. For each site within $20$ms fiber RTT of a \azure{} DC, we compute the $x^{th}$ percentile of its generation time series, cap output at that percentile, and sum capped power across all sites. Fig.~\ref{fig:avail-vs-prov} shows the resulting tradeoff for $3$ geographically diverse \azure{} DCs in the EU spanning $15$ degrees of latitude: Sweden ($60$\textdegree{}N), Netherlands ($52$\textdegree{}N), and Italy ($45$\textdegree{}N).

At the $20$th percentile, the fleet stays above $70\%$ of provisioned power $87$-$89\%$ of the time; rare dips are absorbed by batteries, grid access, and the cross-site router. Lower percentiles
(P$10$) yield higher availability ($>97\%$) with less capacity; higher ones (P$30$) unlock more compute at reduced guarantees. Since guardrails bridge infrequent shortfalls, average GPU utilization
matches the $70$-$90\%$ reported in traditional data centers~\cite{dcsmi_utilization_2024, mckinsey_hyperscaler_2025}. 
At P$20$ \greenf{} could already enroll today $10+$ million H$100$ equivalents.

Having established economic viability and power availability, the remaining challenge is serving AI inference under residual power variability. This motivates a cross-site inference request router that uses power predictions and spatial complementarity and is reactive to workload variability.


\section{AI Inference Under Constrained Power}
\label{sec:background}

Unlike traditional data centers, \greenf{} operates under variable power availability and must carefully navigate the tradeoff between constrained power and AI inference performance.
This section characterizes the action space and its workload impact, informing the router design.



\subsection{\greenf{} Site-local Knobs} 
\label{subsec:llm-inference}

Large language model inference is autoregressive: an input query is absorbed during the \textit{prefill} phase, after which the model generates one token at a time autoregressively in the \textit{decode} phase. The KV-cache stores key-value tensors for previously processed tokens, enabling reuse during generation.  The KV-cache occupancy increases with the number of in-flight tokens and sequence lengths.

An inference system under a power budget can adjust two primary knobs at each site\footnote{We assume homogeneous GPU clusters in each site. Heterogeneous clusters could be treated as multiple homogeneous clusters.}: ($1$)\textit{Active node count}: nodes can be idled ($\sim${}$30\%$ static power tax, instant readiness) or shut down (saves power but minutes of boot latency). Fewer active nodes increases load on remaining instances, queues requests, inflates E$2$E latency, and shrinks the aggregate KV-cache pool, raising memory pressure risk. ($2$)\textit{GPU frequency}: adjustable in milliseconds via \texttt{nvidia-smi}. Lower frequencies reduce throughput, increase TBT, and raise KV-cache occupancy, which, as shown in \S\ref{subsec:gpu-characterization}, can trigger sharp latency degradation beyond a frequency-dependent tipping point. A third knob, tensor parallelism (TP) degree, takes seconds to minutes to reconfigure; we fix TP at deployment to avoid re-sharding overhead. In a multi-site \greenf{} setting with variable renewable power, the challenge is to dynamically select the right <node count, frequency> configuration at each site while distributing requests across sites to meet latency SLOs, an online resource allocation problem. There is still a fourth knob, power-capping, which can be set using \texttt{-pl} in \texttt{nvidia-smi}.

\subsection{Characterizing LLM Inference}
\label{subsec:gpu-characterization}

Designing a robust control plane requires a precise understanding of the underlying hardware's power and performance dynamics. Toward this, we hosted the Llama$3.1$-$8$B model on two NVIDIA A$100$ ($40$~GB) GPUs with tensor parallelism (TP$2$) using vLLM~v$1$~\cite{vllm} serving engine (similar results for H$100$ GPUs are shown below, Fig.~\ref{fig:pp_results}). To enable detailed performance analysis, we modified the vLLM code to log latency metrics outside the critical path. The input workloads were constructed using prefill (P) and decode (D) values from the \azure{} coding and conversation datasets~\cite{azure_trace_dynamollm} each with a distinct P/D ratio. Fig.~\ref{fig:gpu-profiling} shows results for the conversation workload; coding exhibits consistent trends and is omitted for brevity. Request arrivals follow Poisson distributions, and each experiment ran for $30$~minutes. In parallel, we collected fine-grained system metrics using DCGMI~\cite{dcgmi} with samples taken every $50$~ms. We evaluated the system across a wide range of RPS values, scaling up until saturation, and under multiple GPU frequency settings. This exercise focuses on power consumption, inference latency, and KV-cache utilization, metrics that directly govern the feasibility of dynamic scaling. We make the following observations:

\begin{figure*}[h]
    \centering
    \begin{subfigure}[t]{0.33\textwidth}
        \centering
        \includegraphics[width=\linewidth]{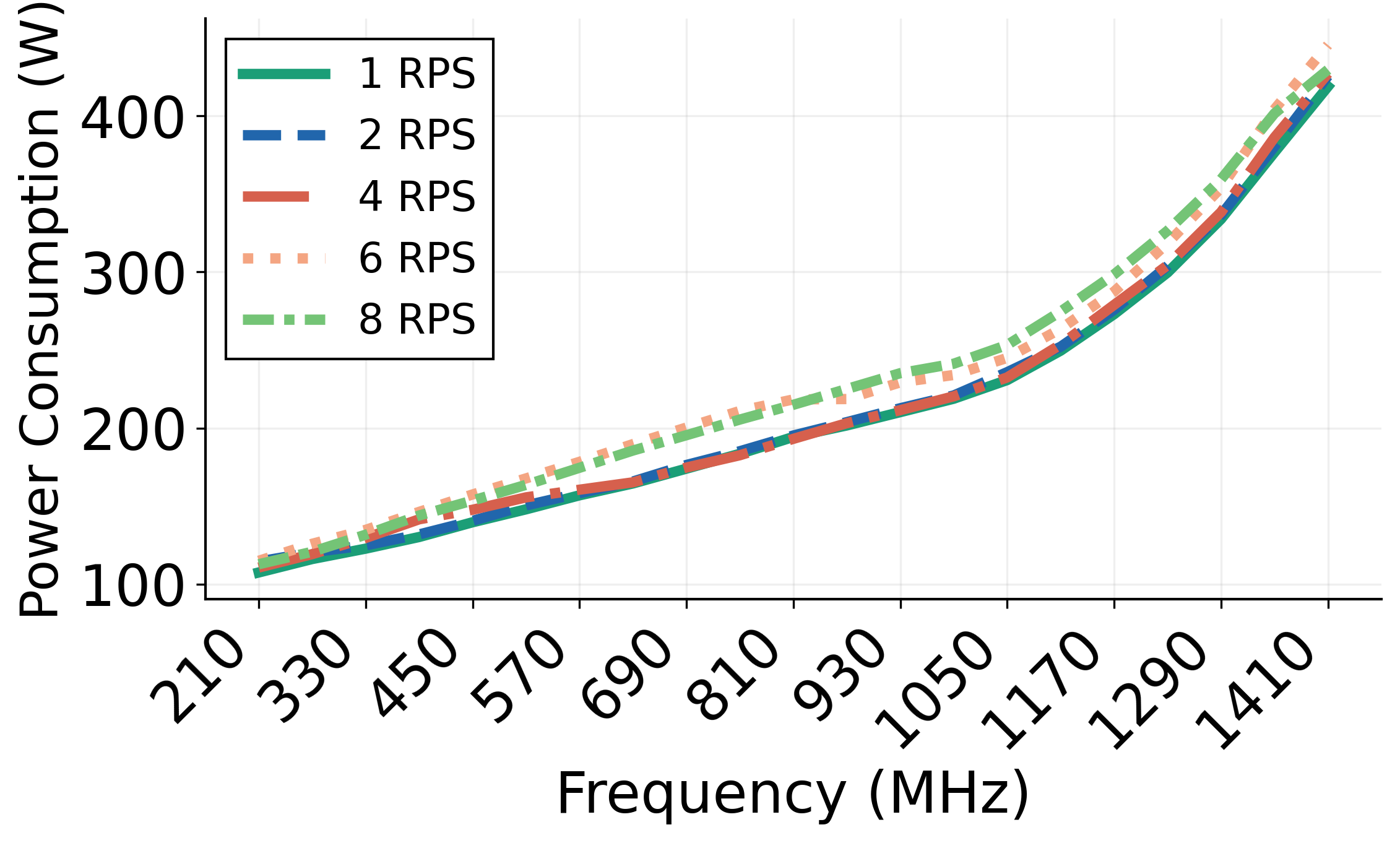}
        \caption{Power vs.\ Frequency ($P99$)}
        \label{fig:power_vs_frequency}
    \end{subfigure}
    \hfill
    \begin{subfigure}[t]{0.33\textwidth}
        \centering
        \includegraphics[width=\linewidth]{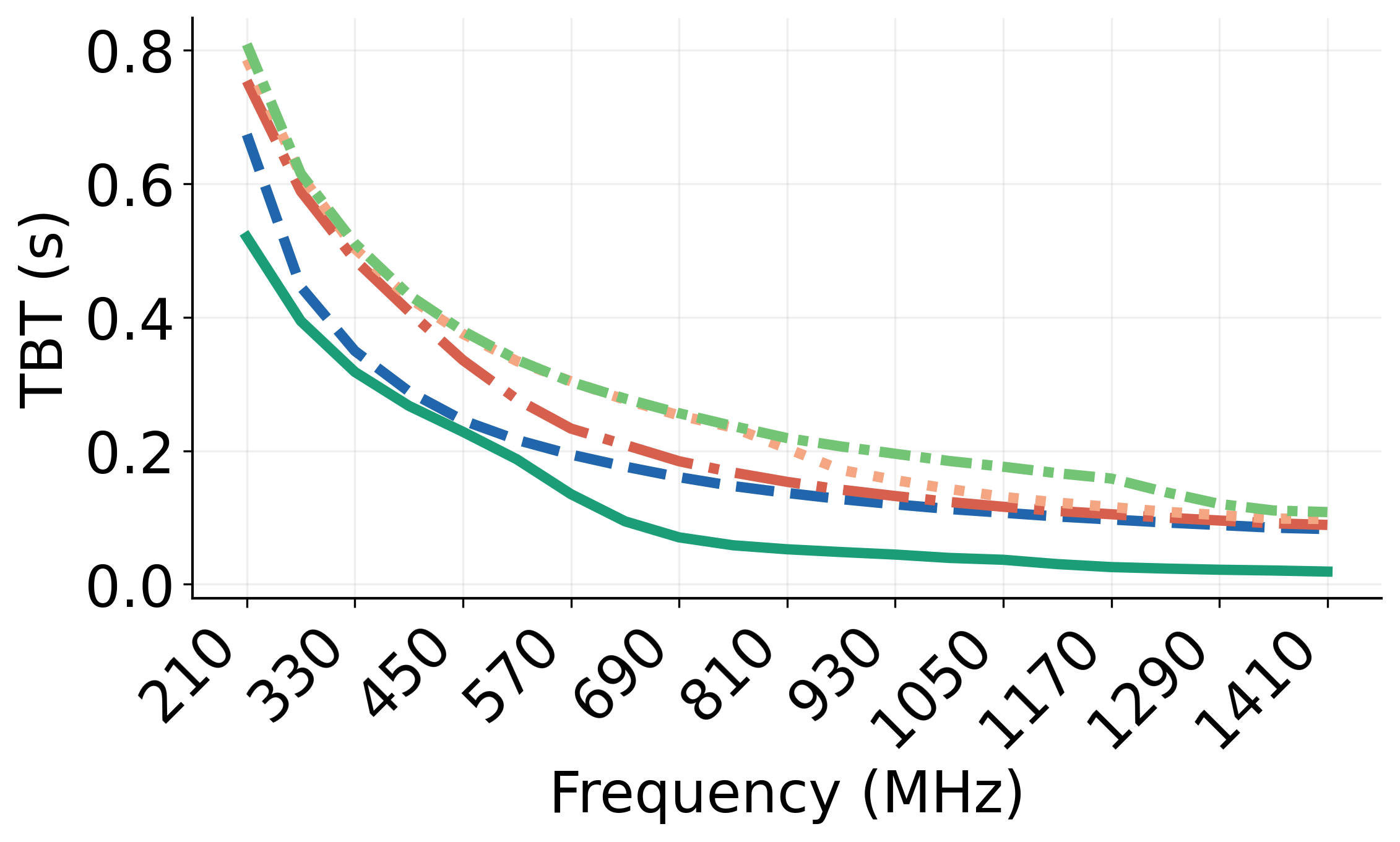}
        \caption{TBT vs.\ Frequency ($P99$)}
        \label{fig:tbt_vs_frequency}
    \end{subfigure}
    \hfill
    \begin{subfigure}[t]{0.33\textwidth}
        \centering
        \includegraphics[width=\linewidth]{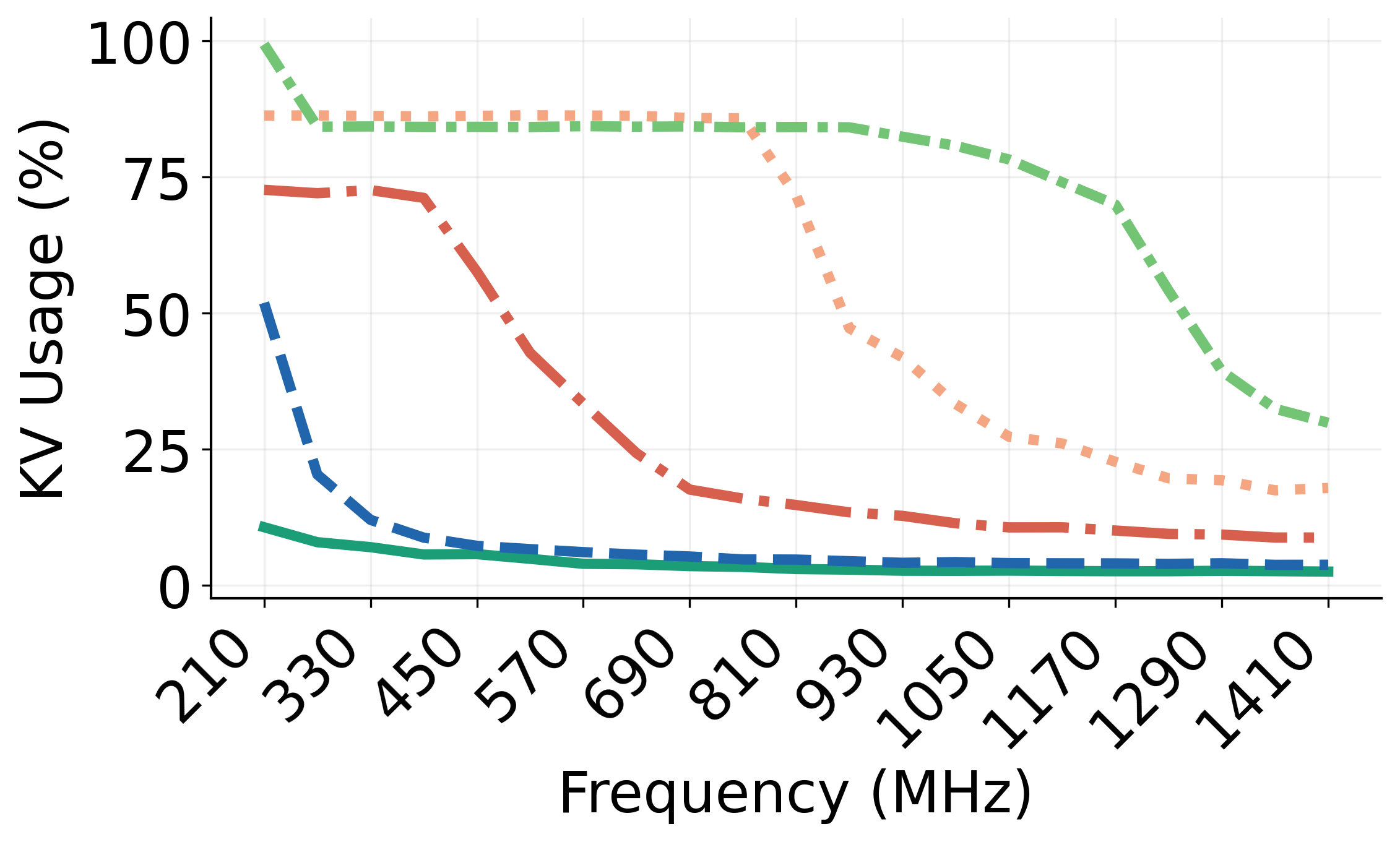}
        \caption{KV Usage vs.\ Frequency ($P99$)}
        \label{fig:kv_vs_frequency}
    \end{subfigure}
    \caption{Profiling results of Llama~$3.1$~$8$B on A$100$ $40$GB for \azure{} conversation workload.}
    \label{fig:gpu-profiling}
\end{figure*}

\parab{O$1$. Frequency versus peak power:} Frequency downclocking is a deterministic knob to restrict power consumption. We stress-tested the system by saturating GPU utilization with high request loads; the power recorded via DCGMI represents peak power at each frequency. A key observation from Fig.\ref{fig:power_vs_frequency} is that the frequency-to-peak-power relationship is non-linear. The lookup table is largely consistent across workloads at low frequencies but exhibits slight RPS-dependence at mid-range frequencies like $810$~MHz. To remain conservative, we build the frequency-to-peak-power lookup table from the peak-load envelope, generated using \texttt{gpu\_burn}~\cite{gpu_burn} at high utilization. 

\greybox{\textbf{Design implication:} \slc{}s use these lookup tables to pick GPU operating frequencies during periods of power constraint.}


\parab{O$2$. Frequency versus inference latency:} Latency scaling with frequency is hardware-dependent. Across $3$ NVIDIA GPU generations (H$100$ $80$GB, A$100$ $80$GB, B$200$), inference latencies (TTFT, TBT, E2E) plateau beyond a frequency threshold due to hardware-enforced power-based throttling; on A$100$ $40$GB GPUs, no such plateau was observed. A common takeaway from Fig.~\ref{fig:tbt_vs_frequency} is that latency does not scale proportionally with frequency. This uncertainty across hardware and frequency-scaling dimensions makes latency an essential additional signal beyond frequency and capacity (active node count). 

\greybox{\textbf{Design implication:} \router{} must use live inference latency signals alongside frequency and capacity signals from individual \greenf{} sites for routing.}

\parab{O$3$. Frequency versus KV-cache usage:} KV-cache usage is the fraction of reserved KV memory occupied by cached key/value tensors. Lowering frequency reduces throughput, so more tokens remain in flight, raising KV-cache occupancy. Since KV-cache size scales with sequence length, larger active contexts increase memory usage and inter-token latency as each generation step attends to all cached tokens. Fig.\ref{fig:kv_vs_frequency} shows a steep increase in KV-usage below a frequency threshold, crucially just before the exponential TBT rise in Fig.\ref{fig:tbt_vs_frequency}. For instance, at $4$RPS, KV-cache pressure intensifies below $690$MHz, serving as a leading indicator of memory bandwidth saturation. Blindly reducing frequency can push the system past this tipping point, triggering massive KV-cache pressure that manifests as sharp TBT degradation. Because this point is workload-dependent, the controller must monitor this signal to tune the safe operating frequency dynamically. Offline profiling identified the optimal KV-cache threshold for the \slc{}: empirically, $20\%$ across all workloads for A$100$ $40$GB GPUs. 

\greybox{\textbf{Design implication:} \slc{}s should avoid downclocking GPUs near this KV-cache threshold, resorting instead to idling some GPUs locally.}

\parab{Generalization to H$100$.} To assess the generalizability of observations O$1$--O$3$, we repeat the same profiling methodology on NVIDIA H$100$ $80$~GB SXM GPUs. Figs.~\ref{fig:h100-power}--\ref{fig:h100-kv} show the frequency-versus-power, -TBT, and -KV-cache results. All three observations hold: (O$1$)~power scales non-linearly with frequency and remains largely workload-invariant; (O$2$)~latency does not scale proportionally with frequency (on H$100$, TBT additionally exhibits a plateau beyond a hardware-enforced throttling threshold, unlike the smooth A$100$ curve); (O$3$)~KV-cache utilization shows a sharp, workload-dependent inflection that precedes TBT degradation. The empirical KV threshold shifts from $20\%$ (A$100$ $40$~GB) to $35\%$ (H$100$ $80$~GB), reflecting the larger KV-cache pool. These results confirm that the \slc{}'s dual-signal design and threshold-based control are broadly applicable across GPU generations; only the threshold values require re-calibration.

\begin{figure*}[t]
    \centering
    \begin{subfigure}[t]{0.33\textwidth}
        \centering
        \includegraphics[width=\linewidth]{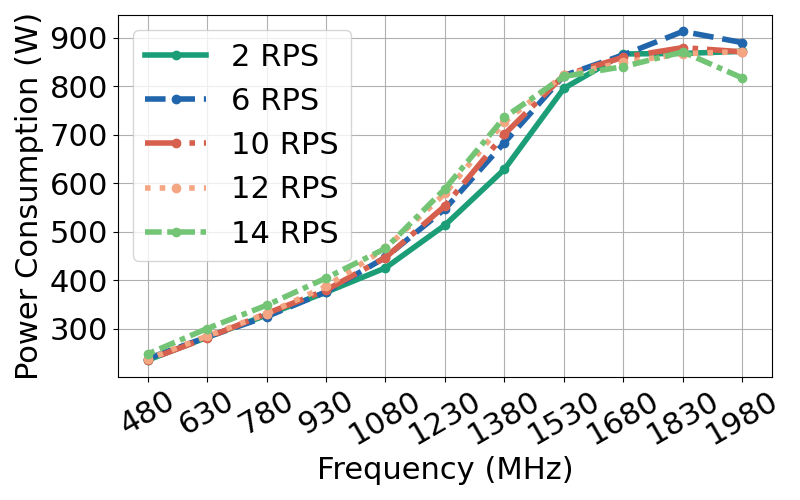}
        \caption{Power consumption vs Frequency (P$99$)}
        \label{fig:h100-power}
    \end{subfigure}
    \hfill
    \begin{subfigure}[t]{0.33\textwidth}
        \centering
        \includegraphics[width=\linewidth]{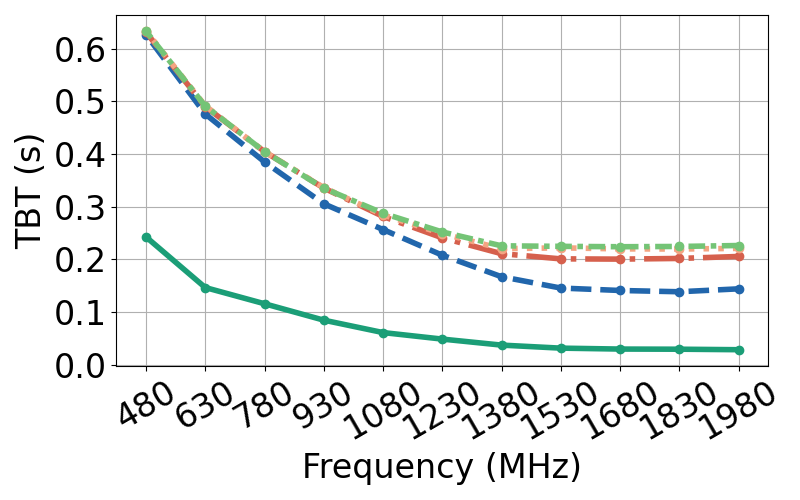}
        \caption{TBT vs Frequency (P$99$)}
        \label{fig:h100-tbt}
    \end{subfigure}
    \hfill
    \begin{subfigure}[t]{0.33\textwidth}
        \centering
        \includegraphics[width=\linewidth]{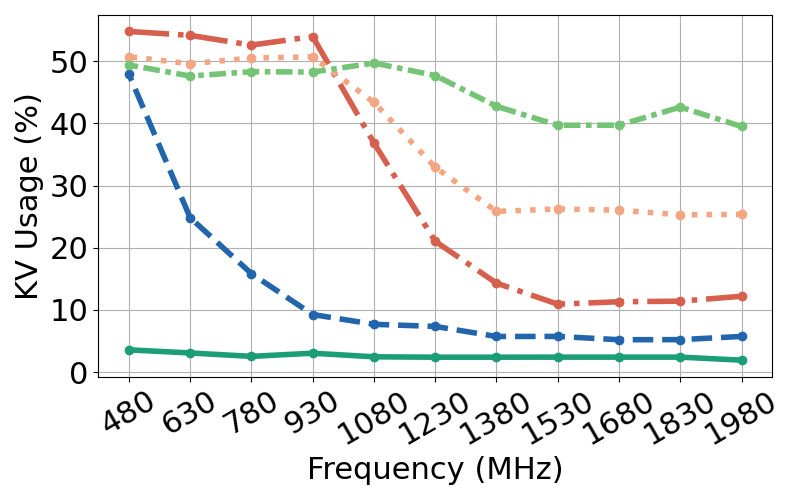}
        \caption{KV Usage vs Frequency (P$99$)}
        \label{fig:h100-kv}
    \end{subfigure}
    \caption{Profiling results of Llama~$3.1$~$8$B on H$100$ $80$~GB for \azure{} conversation workload.}
    \label{fig:pp_results}
\end{figure*}

\section{\router{} Design}
\label{sec:xwind}


We need a robust power-variability-aware request routing and site configuration framework to efficiently route inference requests across multiple \greenf{} sites with temporal variance in power generation while still being able to offer low latencies. 
We propose a lean, production-friendly router, \router{}, that works in tandem with \slc{}s, which reconfigure individual sites and send live capacity and telemetry signals to \router{} router.
Importantly, the \slc{} operates \textit{proactively with respect to power}: it receives a forecasted power budget derived from short-term wind power predictions, enabling it to reconfigure GPU capacity and frequency in anticipation of power changes. The reactive telemetry signals (KV-cache utilization, queue depth, and TBT latency) then serve as \textit{corrective feedback}, refining the configuration within each forecast window.

\subsection{Overview}
\label{subsec:hierarchical-design}

Our high-level design goals for the \greenf{} system are twofold: 
($1$) ensuring graceful operation under residual power variability at individual sites, while
($2$) keeping AI inference latencies low.

A na\"{i}ve routing approach that works with static routing weights (\S\ref{subsec:eval-ablation}) fails in this setting as power availability changes dynamically, leading to significant latency inflation. We rather rely on a site-local \slc{} for local reconfigurations and a cross-site \router{} for routing under these uncertainties with high-level signals from each site. A hierarchical design is necessary for a lean yet robust design: while the \slc{} deals with  hyper-local signals such as KV-cache utilization and queue depth, the \router{} router has a global view of all \greenf{} sites in a region.





\greenf{} sites differ from traditional data centers as they rely on variable power supplies, which requires intelligent reconfiguration of the compute capacities to match the available power. Placing the reconfiguration logic within the \slc{} allows the system to resolve local constraints autonomously. Such a design ensures that while the \slc{} handles high-frequency hardware adjustments locally, the cross-site router focuses on global visibility and coarse-grained signals re-balancing inferencing traffic as necessary.

\parab{Why not an optimal/centralized approach?} 
\router{} is deliberately suboptimal: 
global optimality would require a centralized approach, offline workload profiling, and accurate predictions of both request arrival rates and response output lengths, none of which are practical in real production deployments without significant overhead. 
A centralized approach also struggles to scale, as it can be easily overwhelmed by the fine-grained, site-level telemetry signals discussed in \S\ref{subsec:slc}, making it ill-suited for production environments.

\parab{Selective engagement.} \greenf{} sites can accommodate batteries, additional renewable modalities, or opportunistic grid draw. Moreover, peak power needs at individual sites are much lower percentiles of the renewable sites' peak generation capacities, and these infrastructure-level measures significantly improve site availability. The software routing layer need only handle rare residual shortfalls after infrastructural leverages are exhausted. Our system remains passive when all sites meet peak demand; only when a \slc{} anticipates a power crunch does it trigger local hardware reconfigurations and signal \router{} for cascading routing adjustments.

\begin{figure}[t]
    \centering
    \includegraphics[width=0.8\columnwidth]{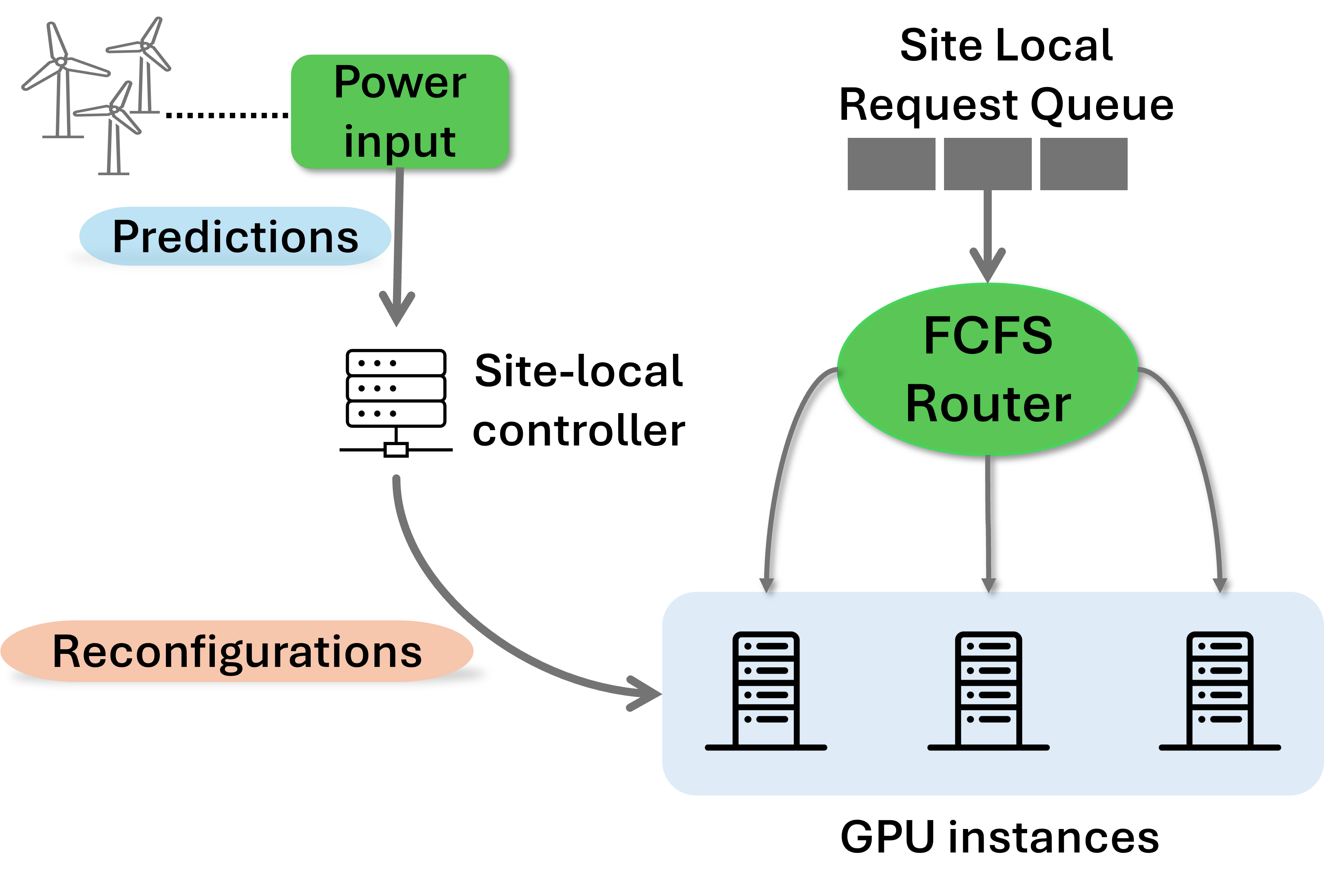}
    \caption{A \greenf{} site of Fig.~\ref{fig:xwind-architecture}.}
    \label{fig:greenferencing-site}
\end{figure}

\subsection{\router{} Site-local Controller (\slc{})}
\label{subsec:slc}

\greenf{} sites must dynamically scale compute capacity to remain within the instantaneous power budget, primarily by adjusting active node count or operating frequency. Conventional approaches tune only a single knob, which is inherently sub-optimal: frequency downclocking alone is inefficient under moderate loads, forcing the entire cluster to suppressed clock speeds when consolidating onto fewer high-frequency nodes would be more effective. Conversely, node shutdown alone creates quantization inefficiencies, stranding usable power when availability drops even marginally. Idling nodes avoids boot latency but incurs a high static power tax ($\sim${}$1$,$920$~W idle per A$100$ DGX, roughly $30\%$ of peak $6$,$500$~W). Our \slc{} design achieves the right balance by jointly exploring the state space of node counts ($N$) and frequencies ($f$), using real-time telemetry to filter configurations that satisfy current demand.

\parab{Decision window.} The \slc{} operates on a $3$-minute decision cycle, yielding five discrete steps per $15$-minute forecast window (with linear interpolation), allowing gradual transitions to the target power state. This interval balances two concerns: shorter cycles prevent \router{} (running every $15$ seconds) from converging on stable weights between reconfigurations, while longer cycles force conservative node idling at window start to accommodate anticipated power drops, stranding usable capacity.

\parab{Candidate generation.} When anticipating a power change, the \slc{} enumerates all feasible ($N$, $f$) tuples satisfying the new constraint, using the workload-invariant peak-power and frequency mappings from profiling (O$1$, \S\ref{subsec:gpu-characterization}). For each candidate node count (1 to $n$), it computes the per-node power budget and the corresponding maximum frequency. Since capping all GPUs to one frequency may strand power, the \slc{} boosts a fraction of GPUs to the next frequency level to fully utilize the power budget.

\parab{Candidate filtering via telemetry.}
Selecting the optimal ($N$, $f$) tuple is challenging because the workload characteristics (input length, output length, and arrival rates) exhibit significant spatiotemporal volatility.
A workload-agnostic strawman approach may select the tuple maximizing $N \times f$, since total FLOPs scale with this product. 
However, our experiments show that sub-maximal products can outperform the theoretical best depending on workload characteristics. 
For instance, under high RPS, using all nodes at a slightly lower frequency may beat using two-thirds of nodes at higher frequency, as the former provides more parallel entry points to mitigate queuing delays.

To tackle this challenge, the \slc{} leverages temporal locality at the macro level. Since aggregate workload trends typically shift at coarser timescales, the \slc{} uses system and latency metrics from the most recent time window as a proxy for near-future demand. It uses the following signals:

\begin{myitemize}
    \item \textbf{Queue depth:} The number of requests waiting to be scheduled. Sustained queue build-up indicates insufficient compute capacity and prevents the \slc{} from reducing the number of active nodes.
    \item \textbf{KV-cache utilization:} The fraction of reserved KV-cache memory occupied by in-flight tokens. High KV utilization signals rising HBM bandwidth demand and an approaching memory-bound regime; when it crosses a threshold, the \slc{} raises the operating frequency to avoid the latency knee point (O$3$, \S\ref{subsec:gpu-characterization}).
    \item \textbf{TBT latency:} The median token-by-token latency across GPUs in the previous window. TBT captures SLO violations and non-memory bottlenecks such as thermal throttling or compute-bound phases.
\end{myitemize}

\parab{The reactive control policy.} Using the telemetry signals described above, we design a reactive heuristic (Algorithm~\ref{alg:slc-a}) that dynamically determines the minimum operating frequency and active node count to maintain stability. 

At a high level, it maintains two state variables, $F_{\text{floor}}$ and $N_{\text{curr}}$, which are updated based on real-time signals. If a queue build-up is detected, $N_{\text{curr}}$ is updated to match the current number of active nodes. If KV-cache utilization exceeds a hardware-dependent threshold, $F_{\text{floor}}$ is increased by $2\Delta f$; similarly, if the TBT latency exceeds its threshold, $F_{\text{floor}}$ is raised by $\Delta f$. The candidate selection process applies these constraints dynamically. In the event of a queue build-up, the \slc{} discards all candidates with a node count lower than $N_{\text{curr}}$ and selects the remaining tuple with the highest $N \times f$. In the absence of queuing, the \slc{} filters out candidates operating below $F_{\text{floor}}$ and selects the configuration that maximizes the capacity product for the next cycle.

\begin{algorithm}[t] 
\caption{Reactive Site-Level Controller (\slc{})}
\label{alg:slc-a}
\begin{algorithmic}[1]

\Statex \textbf{Input:} Power budget $P_t$; telemetry $\phi_t = (\text{KV}_t, Q_t, L_t)$
\Statex \textbf{State:} Active GPUs $N_{\text{curr}}$, frequency $f_{\text{curr}}$, floor $F_{\text{floor}}$

\Function{Reactive\slc{}}{$P_t,\;\phi_t$}

    \State $\mathcal{T} \gets \{(f, N) : \text{Power}(f,N) \le P_t\}$ \Comment{Viable Configs}
    \State $\pi \gets (Q_t > Q_{\text{max}})$ \Comment{Congestion flag}

    \Statex \hspace{1em} \textit{// Adjust frequency floor from telemetry}
    \If{\textbf{not} $\pi$}
        \If{$\text{KV}_t > \text{KV}_{\text{max}}$} $F_{\text{floor}} \mathrel{{+}{=}} 2\Delta_f$
        \ElsIf{$L_t > L_{\text{max}}$} $F_{\text{floor}} \mathrel{{+}{=}} \Delta_f$
        \ElsIf{$L_t < L_{\text{max}} \wedge \text{KV}_t < \text{KV}_{\text{max}} \wedge Q_t < Q_{\text{max}}/2$}
            \State $F_{\text{floor}} \mathrel{{-}{=}} \Delta_f$
        \EndIf
        \State $F_{\text{floor}} \gets \text{clamp}(F_{\text{floor}},\; F_{\text{min}},\; F_{\text{max}})$
    \EndIf

    \Statex \hspace{1em} \textit{// Select best $(f, N)$ maximizing $f \cdot N$}
    \State $\mathcal{S} \gets \begin{cases} \{(f,N) \in \mathcal{T} : N \ge N_{\text{curr}}\} & \text{if } \pi \\ \{(f,N) \in \mathcal{T} : f \ge F_{\text{floor}}\} & \text{otherwise} \end{cases}$
    \State \Return $\arg\max_{(f,N)\,\in\, \mathcal{S} \cup_\emptyset \mathcal{T}} f \cdot N$

\EndFunction
\end{algorithmic}
\end{algorithm}


\subheading{Capacity priority over frequency.} When queues build up, we prioritize a higher $N$ over $f$. Frequency scaling yields diminishing returns in memory-bound LLM inference, whereas adding a node scales both throughput and aggregate KV-cache capacity linearly. Reducing node count under high load causes queuing delays to inflate exponentially as new arrivals face reduced service capacity. Prioritizing $N$ maintains a larger memory pool and sufficient concurrency to keep the queue stable.


\subheading{Dual signals for $F_{\text{floor}}$.} We employ two orthogonal signals. KV-cache utilization is a \textit{leading indicator} of memory bandwidth saturation (O$3$): the \slc{} raises $F_{\text{floor}}$ on KV-cache pressure, preemptively defending against memory bottlenecks before they surface as user-facing latency. TBT is a \textit{lagging indicator} that captures non-memory bottlenecks, such as thermal throttling or compute-bound prefill phases, which KV-cache alone would miss.

\subheading{Asymmetric correction steps.} The frequency correction applies differential gain: $2\Delta f$ for KV vs.\ $\Delta f$ for TBT. KV-cache usage exhibits a sharp saturation cliff at lower frequencies, necessitating aggressive correction to immediately exit saturation and prevent out-of-memory errors. TBT follows a smooth degradation curve, permitting finer-grained adjustments that avoid overshooting.

\subheading{Maximizing live capacity.} After filtering for stability, the \slc{} selects the tuple that maximizes $N \times f$, ensuring the system fully utilizes the available power budget and eliminates stranded power.

\subsection{\router{} Cross-Site Router}
\label{subsec:xwind-router}

A \router{} cross-site router sits in a data center region and oversees multiple \greenf{} sites.
Its main responsibility is to distribute the incoming inference requests across the associated sites.
A strawman approach might use node count as a static weight for round-robin balancing, but at green energy sites, compute capacity fluctuates with the underlying power source, making fixed weights inadequate.
\fix\router{} therefore updates routing weights every second along two paths: ($1$) the \emph{proactive} path reacts immediately to capacity or frequency changes signaled by the \slc{}s and ($2$) the \emph{reactive} path that corrects residual latency imbalance using observed TBT. Algorithm~\ref{alg:xwind} summarizes the procedure; the rest of this subsection walks through its components.

\parab{Routing metric: live compute capacity.} 
\router{} operates on live compute capacity. It periodically probes each \slc{} for its current active configuration ($N$, $f$). The routing weight $W_i$ for site $i$ is: $W_i = N_i \times f_i$.
This product serves as a robust proxy for the site's instantaneous token-processing capability, as FLOPs scale linearly with operating frequency~\cite{flops_and_frequency}.

\parab{The latency-corrective feedback loop.} Relying solely on $N \times f$ can cause imbalances due to non-linear scaling (O$2$, \S\ref{subsec:gpu-characterization}). \router{} therefore implements a corrective feedback loop, polling \slc{}s every second for active node count, frequency, and TBT. On capacity or frequency changes, weights immediately reset proportionally to the new $N \times f$. When no change occurs for $15$~seconds, a latency-corrective loop computes an exponential moving average of each site's TBT, calculates the ratio to the global mean, clips it within a sensitivity bound $\delta$, and asymmetrically penalizes only above-mean sites to prevent oscillatory migration. The weights are then renormalized. The \slc{} signals \router{} $5$~seconds before any change in node-count to prevent transient imbalances.

\parab{Breaking the cycle.} The \slc{}'s dual-knob design introduces a potential oscillation cycle: KV-cache rises $\rightarrow$ frequency floor raised $\rightarrow$ node idled to stay in power budget $\rightarrow$ queue builds up $\rightarrow$ capacity-priority restores node count at lower frequency $\rightarrow$ KV-cache rises again. \router{}'s cross-site load redistribution breaks this cycle: when the \slc{} idles a node, \router{} immediately recalculates routing weights ($w_i = N_i \times f_i$), reducing traffic in proportion to diminished capacity. The arrival rate drops in lock-step with the node reduction, preventing the queue build-up that would trigger the next oscillation step. Additional safeguards include asymmetric hysteresis (frequency floor decreases require all three signals to be simultaneously benign) and a capacity-priority circuit breaker (once the queue exceeds $Q_{\text{max}}$, the controller locks node count and only optimizes frequency).
\begin{algorithm}[H]
    \caption{\router{} Adaptive Weight Update (every $1$\,s)}
    \label{alg:xwind}
    \begin{algorithmic}[1]
        \Statex \textbf{Input:} $\ell, c, f$ -- latency, capacity, frequency vectors over sites $\mathcal{S}$
        \Statex \textbf{State:} $w_s, \hat{L}_s$ -- weight and EMA-smoothed latency per site
        \Statex \textbf{Params:} $\alpha$ (EMA factor), $\delta$ (sensitivity bound), $\Delta t{=}15$\,s


        \If{$c \neq c_{\text{prev}}$ or $f \neq f_{\text{prev}}$ } \Comment{Proactive path}
            \ForAll{$s \in \mathcal{S}$}
                \State $w_s \gets \frac{c_s \cdot f_s}{\sum_{s} c_{s\_prev} \cdot f_{s\_prev}} \cdot \sum_{s} c_{s}$
            \EndFor
        \ElsIf{$\Delta t$ elapsed} \Comment{Reactive path}
            \ForAll{$s \in \mathcal{S}$}
                \State $\hat{L}_s \gets (1{-}\alpha)\hat{L}_s + \alpha\,\ell_s$
            \EndFor
            \State $\bar{L} \gets \text{mean}(\hat{L})$
            \ForAll{$s \in \mathcal{S}$}
                \State $\rho_s \gets \text{clip}(\hat{L}_s/\bar{L},\; 1{-}\delta,\; 1{+}\delta)$
                \State $w_s \gets w_s / \rho_s$ \textbf{if} $\rho_s > 1$ \Comment{Reduce only slower sites}
            \EndFor
            \State Normalize: $w_s \gets w_s \cdot \sum c_{s} / \sum w_{s}\ \forall s$
        \EndIf
    \end{algorithmic}
\end{algorithm}

\section{Evaluation}
\label{sec:evaluation}

Our evaluation focuses on the following questions:
\vspace{0.1in}
\begin{myenum}
    \item[\textbf{Q$1$}] How does \router{} compare against baselines under cross-site power-constrained operations across workload types and load levels? 
     \item[\textbf{Q$2$}] Why does the reactive \slc{} outperform Max-FLOPS, and 
     when does it choose to diverge?
     \item[\textbf{Q$3$}] Are both control signals (KV-cache and TBT) necessary for the \slc{} decisions?
     \item[\textbf{Q$4$}] How much does each cross-site routing signal contribute 
     to \router{}'s end-to-end performance?
\end{myenum}

\subsection{Experimental Methodology}
\label{subsec:eval-methodology}

We emulate a multi-site \greenf{} deployment on a testbed of $64$ NVIDIA A$100$ $40$~GB GPUs across three sites (Site-$0$, Site-$1$, Site-$2$) with a $2$:$1$:$1$ allocation ($32$, $16$, $16$ GPUs). Each GPU pair hosts one Llama$3.1$ $8$B~\cite{llama3} instance served via vLLM~\cite{vllm} with TP$2$, yielding $16$, $8$, and $8$ model instances respectively. We modified vLLM to asynchronously log request-level (E$2$E, TBT, TTFT, queuing delay) and system-level (queue length, KV-cache utilization) metrics. A lightweight Instance Telemetry process computes sliding-window averages (window=$15$~s, step=$1$~s) and exposes state to the \slc{}, which polls every $15$~s to construct site-wide aggregations for \router{}. The \slc{} also schedules (round-robin) requests across homogeneous local instances and signals \router{} five seconds before any change in active node count.

\parab{Request arrival traces.} We generate $120$-minute synthetic traces using Poisson arrivals, sampling prefill and decode lengths from \azure{} production traces. We evaluate three workload types: \textit{coding} (prefill-to-decode ratio $114$, prefill-heavy), \textit{conversation} (ratio $16$, decode-heavy), and \textit{mixed} (uniform random selection from both). Prompt tokens are generated using the Llama~$3.1$ $8B$ tokenizer. Our custom load generator replays traces with an internal timer synchronized to the trace timeline, keeping arrival rates independent of system latency: critical for evaluating queue dynamics during power contention. We test for $150$~RPS (requests per second; moderate) and $175$~RPS (high load).

\begin{figure}[t]
    \centering
    \includegraphics[width=0.8\columnwidth]{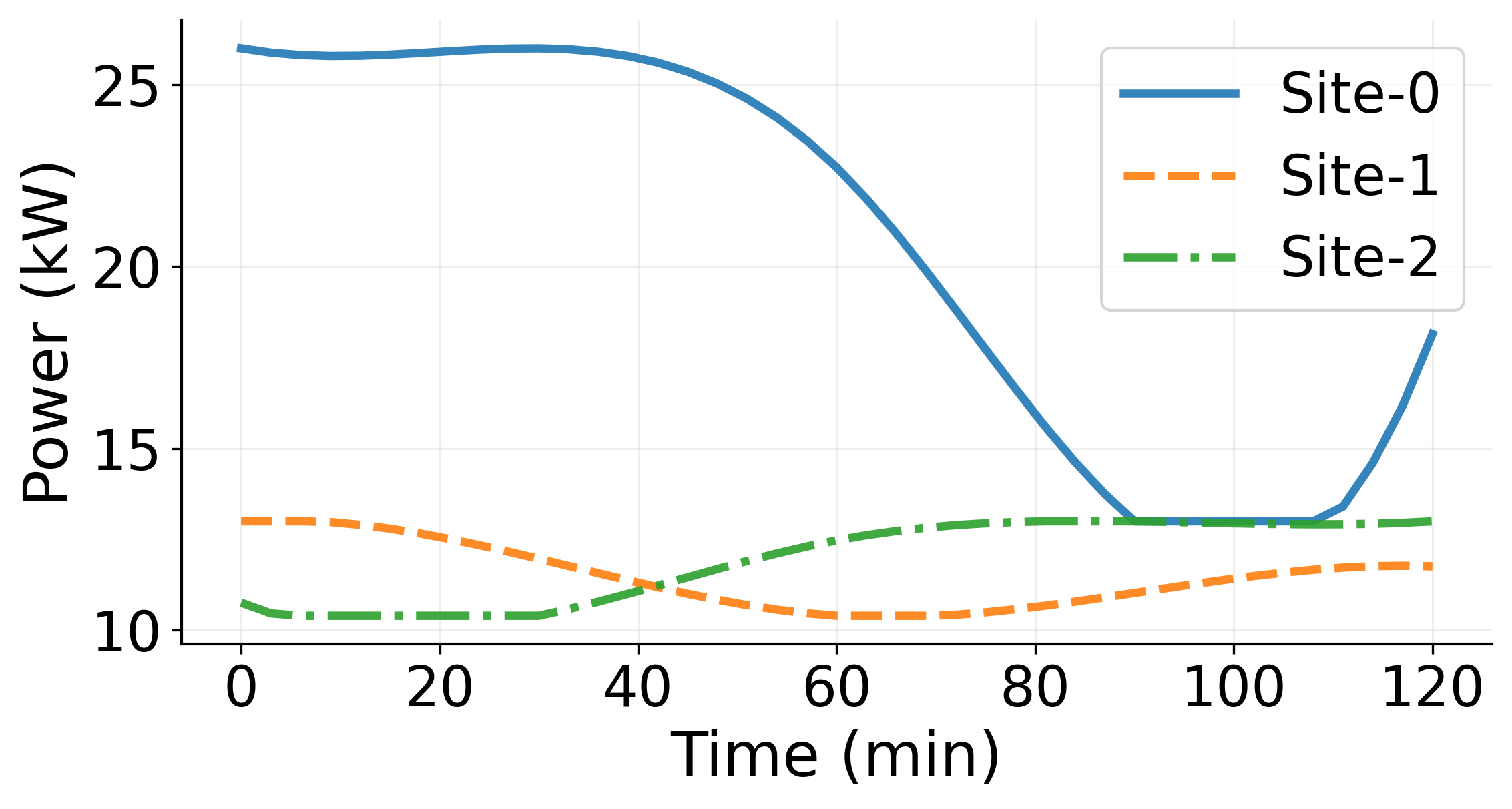}
    \caption{Power availability across $3$ sites scaled from real US wind farm data. Site-$0$ experiences a sustained $\sim${}$50\%$ power drop mid-trace. Note: $y$-axis does not start at $0$.}
    \label{fig:power-profile}
\end{figure}

\parab{Power trace.} We derive our power profiles from real-world wind power data, scaled to match the power footprint of our GPU testbed. We select three wind farm sites in the Central US that exhibit spatial complementarity in generation. The $2$-hour power trace (Fig.~\ref{fig:power-profile}) captures a realistic scenario where Site-$0$ (the largest site) experiences a sustained power drop of approximately $50\%$ mid-trace, while one of the smaller sites dips by $20\%$. In our experiments, we deliberately do not use a backup data center to stress test the \greenf{} deployment. All requests are served by the three wind-powered sites; none are dropped or rejected. 

\parab{Power forecasting.} As mentioned in \S\ref{subsec:taming-uncertainty}: here we assume oracular knowledge of power availability at $15$~minute intervals and linearly interpolate between them, providing the \slc{} with a continuous power budget for each $3$-minute decision cycle. Enough redundancy in the power distribution and uninterruptible power supply units tackle any residual inaccuracies in prediction and also offers smoothened rather than abrupt changes in the input. Finally, empirically across GPU generations, workloads, and power profiles, a longer decision cycle wastes available power, while a shorter cycle results in convergence challenges at the \router{} router.

\parab{GPU power modeling.} GPU servers also incur non-GPU power overhead (CPUs, NICs, cooling). We amortize this by assigning a fixed $240$~W ($1$,$920$~W idle state consumption split uniformly across $8$~GPUs in a DGX) per active A$100$ in our power consumption models. This value could be trivially changed, as needed (we tested for H$100$s with a different assigned value), without affecting the utility of our solution.



\parab{Network latency.} \greenf{} sites are co-located within a region; so network latency inflation is minimal and only affects the time when the user receives back the first token (not time between following tokens that are streamed). The marginal TTFT inflation is well within typical user-facing TTFT SLO bounds (few $100$~ms to seconds), making it orthogonal to the power-driven latency dynamics that are the focus of our evaluation. Hence, we can safely ignore this latency component in our emulations.

\subsection{Baselines and Metrics}
\label{subsec:baselines}

We compare our reactive \slc{} with four approaches, each paired with \router{} for cross-site balance. ($1$) \textit{Downclock}: computes per-node power budgets and locks GPUs to the maximum feasible frequency via the workload-invariant lookup table (O$1$). ($2$) \textit{Idle}: inactive nodes stay at lowest frequency with no requests ($\sim${}$30\%$ peak power tax), enabling instant ramp-up; active instances run at max. $1$,$410$ MHz. ($3$) \textit{Power-Capping}: applies hardware power limits via \texttt{nvidia-smi -pl}; GPU frequency could vary with workload, so \router{} uses observed aggregate frequency over a sliding window for routing weights. ($4$) \textit{Max-FLOPS}: maximizes $N \times f$ within the power budget: an approach equivalent to reactive \slc{} but without telemetry-driven filtering.


\parab{Metrics.} Since no requests are dropped, the goal is graceful degradation under power contraction. We report: ($1$)\textit{P$99$ E$2$E latency}, capturing the tail user experience, including queuing and inference time, and ($2$)\textit{P$99$ queue time}, isolating the waiting time before inference. Since all approaches use identical hardware and model configurations, inference time at a given frequency is roughly constant across baselines. The differences in E$2$E latency are driven by queuing, making queue time P$99$ the diagnostic metric for routing and capacity decisions.

\parab{Thresholds.} \slc{} thresholds are calibrated from offline A$100$ $40$~GB profiling (\S\ref{subsec:gpu-characterization}). \textit{KV-cache threshold} $\text{KV}_{\max}=20\%$, set at the sharp inflection in P$99$ KV utilization versus frequency (Fig.\ref{fig:kv_vs_frequency}). \textit{TBT threshold} $L_{\max}=100$~ms per typical interactive SLOs~\cite{dynamollm}. Asymmetric corrections ($2\Delta f$ for KV vs. $\Delta f$ for TBT) reflect the steeper KV cliff versus gradual TBT curve; $\Delta f=60$~MHz (one A$100$ step). \textit{Queue depth threshold} $Q_{\max}=5$ per instance; exceeding this triggers capacity-priority mode (\S\ref{sec:xwind}). Decision cycle $T_{\text{cycle}}=180$~s balances queue drain time against $15$-minute power forecast granularity. \router{} probes \slc{}s every $1$~s and rebalances routing weights every $15$~s.

\subsection{\slc{} performance}
\label{subsec:eval-results}

\begin{figure*}[t]
    \centering
    \includegraphics[width=.9\textwidth]{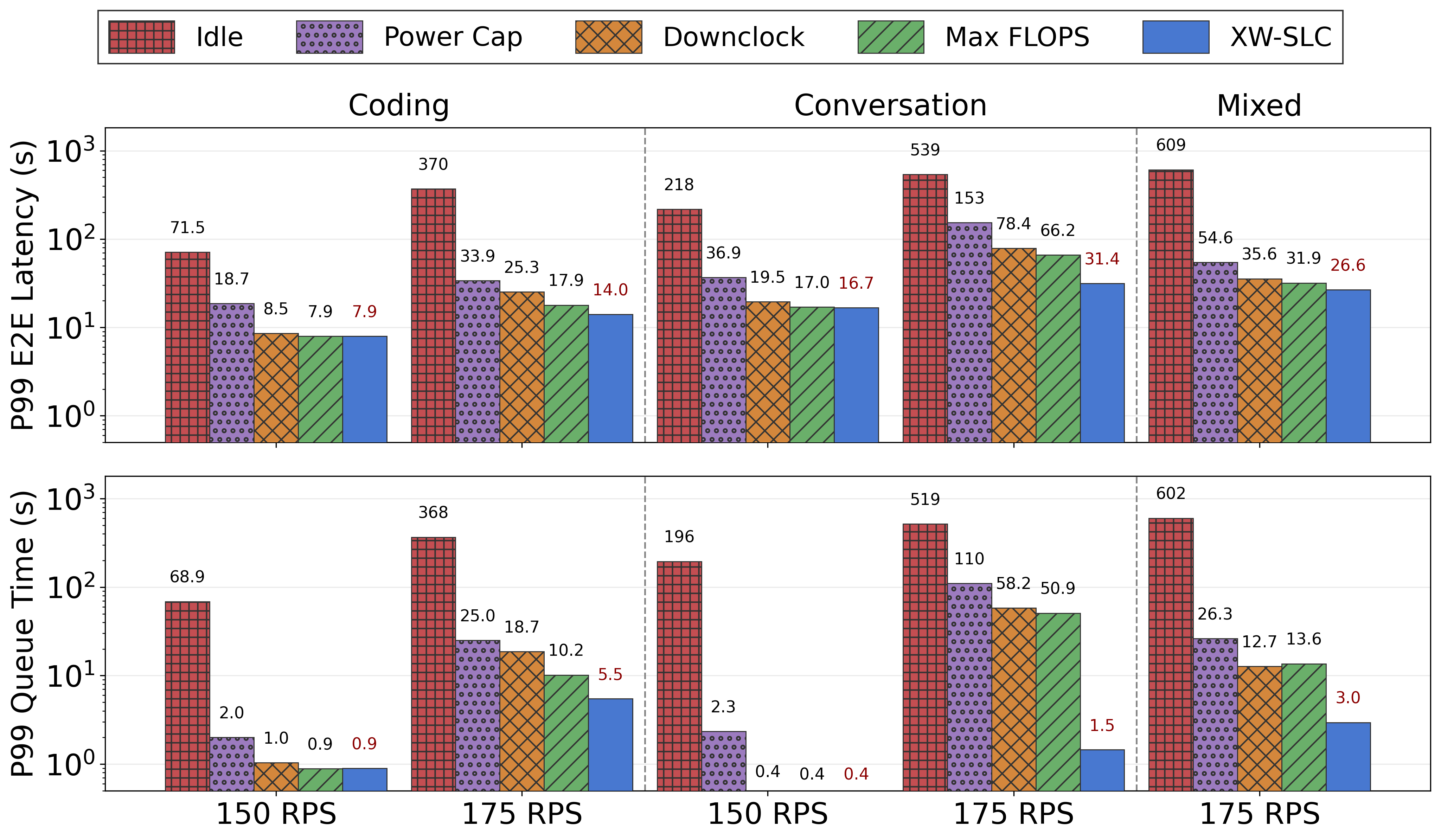}
    \caption{P$99$ E$2$E latency (top) and P$99$ queue time (bottom) across site-local logic variants for coding and conversation workloads. Note: $y$-axis is log-scale.} 
    \label{fig:results}
\end{figure*}

Fig.~\ref{fig:results} shows P$99$ E$2$E latencies and P$99$ queue times across all site-local baselines and \slc{}, with \router{} as the cross-site router, for three workload types at $150$ and $175$ RPS.

\parab{No single knob suffices.} 
Single-knob baselines degrade catastrophically under power contraction.
At $175$~RPS, \textit{Idle} reaches $370$~s (likewise, $539$~s) for coding (conversation).
Essentially, the static power tax of idle nodes wastes desperately needed capacity.
\textit{Power-Capping} reaches $33.9$~s ($153$~s): its non-deterministic frequency behavior causes unpredictable throughput collapses during power drops. Even \textit{Downclock}, the most competitive single-knob approach, reaches $25.3~s$ ($78.4$~s) at $175$~RPS: $1.8\times$ ($2.2\times$) worse than \slc{}. The queue-time breakdown (Fig.~\ref{fig:results}, bottom) shows that these E$2$E gaps are driven primarily by queuing: \textit{Idle} accumulates $368$~s of queue time at $175$~RPS, whereas \slc{} keeps queuing to $5.5$-$27.4$~s. These results validate the need for a dual-knob site-local strategy, as in \slc{} (\textbf{Q$1$}), that simultaneously adjusts frequency and active node count.

\parab{\slc{} versus Max-FLOPS.} \textit{Max-FLOPS}, while devoid of the telemetry-driven filtering in \slc{}, also operates both knobs. While at $150$~RPS, the performances are still comparable, at $175$~RPS \slc{} offers $22\%$ (likewise, $52\%$) lower P$99$ E$2$E latencies for coding (conversation) traces. A deeper investigation into the decisions taken by the schemes at each step revealed that \slc{} could more aggressively trade parallelism (active node count) for frequency than \textit{Max-FLOPS}. This demonstrates that telemetry filtering (queue depth, KV-cache, TBT) is important for an efficient \slc{} design in a variable-power setting (answers \textbf{Q$2$}).

\begin{figure}[tbh]
    \centering
    \includegraphics[width=0.8\columnwidth]{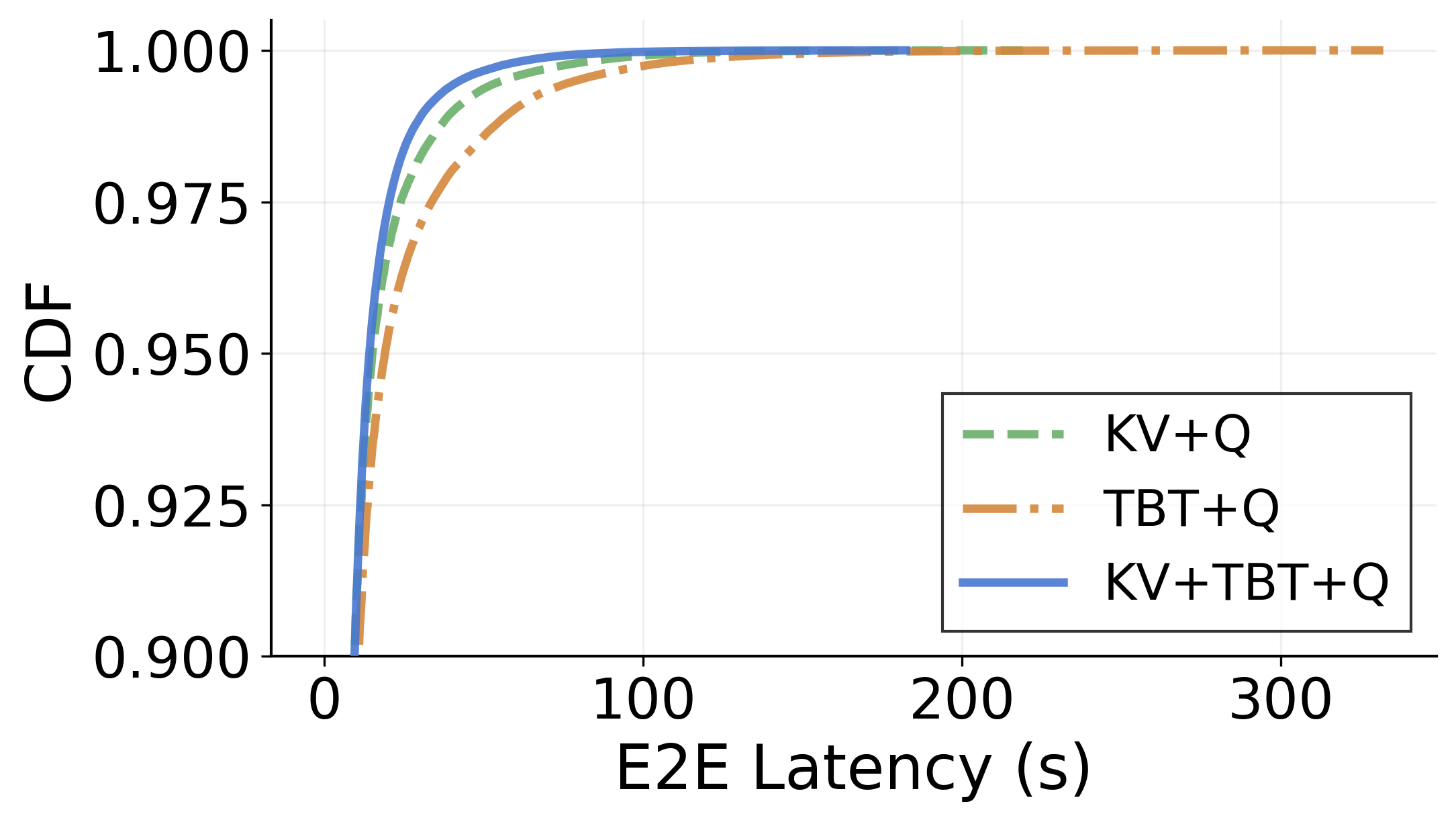}
    \caption{CDF (tail) of E$2$E latency for conversation trace, $175$~RPS, for different site-local logic.} 
    \label{fig:slc_ablation}
\end{figure}

\parab{\slc{}'s Dual-signal for frequency fine-tuning.} We validate the dual-signal design by running conversation at $175$~RPS with three \slc{} variants: full (KV+TBT+Q; Q being the queue depth), KV-only (KV+Q), and TBT-only (TBT+Q). Fig.\ref{fig:slc_ablation} shows tail E$2$E latency CDFs. Removing TBT has modest impact (P$99.9$ shifts from $\sim${}$72$~s to $\sim${}$96$~s), consistent with KV being the dominant signal. Removing KV causes severe degradation (P$99.9$ exceeds $120$~s): without this leading indicator, the \slc{} reacts only after TBT spikes, at which point KV saturation cascades into sustained queuing. The full system achieves the tightest tail: KV provides early warning against memory-driven blow-up, while TBT backstops non-memory bottlenecks (thermal throttling, prefill surges, hardware power-limit frequency drops) that degrade latency without manifesting in KV
occupancy (answers \textbf{Q$3$}).


\parab{\slc{} deep-dive on the conversation trace.} 
At $175$~RPS, \slc{} diverges from \textit{Max-FLOPS}.
During power contraction at Site-$0$ ($t=5$,$400$–$7$,$200$~s), KV utilization repeatedly exceeds $20\%$, driving the dynamic frequency floor from $600$ to $840$~MHz. This frequency-over-parallelism trade-off exploits a property of decode-heavy requests (P/D=$16$): per-token latency scales directly with GPU frequency, compounding over hundreds of tokens. \textit{Max-FLOPS}'s $14$ instances at $540$~MHz offer more parallel slots, but each drains slowly, accumulating KV entries until memory bandwidth saturates. 
Fewer slots at higher frequency ($840$ vs. $540$~MHz) drain requests faster, keeping KV utilization in check and preventing cascading TBT degradation. The \slc{} discovers this trade-off at runtime purely from the KV signal, without workload labels.


\parab{Mixed workload.} To confirm that \slc{}'s gains are substantial even for a mixed workload, we repeat the same experiment but the prefill and decode lengths are sampled uniformly at random from both coding and conversation distributions. At $175$~RPS (Fig.~\ref{fig:results}, right), \slc{} offers a $17\%$ reduction in P$99$ E$2$E compared to \textit{Max-FLOPS}. All other baselines inflate latency even more.
\begin{table}[b]
\caption{P$95$/P$99$/P$99.9$ E$2$E latency (in seconds) for all site-local methods for coding and conversation workloads at $150$ and $175$~RPS. \textbf{Bold} highlights \slc{}.}
\label{tab:percentiles}
\centering
\small
\setlength{\tabcolsep}{4pt}
\begin{tabular}{@{}l rrr rrr@{}}
\toprule
 & \multicolumn{3}{c}{$150$ RPS} & \multicolumn{3}{c}{$175$ RPS} \\
\cmidrule(lr){2-4} \cmidrule(lr){5-7}
Site-local Logic & P$95$ & P$99$ & P$99.9$ & P$95$ & P$99$ & P$99.9$ \\
\midrule
\multicolumn{7}{@{}l}{\textit{Coding Workload}} \\[2pt]
\textbf{\slc{}} & \textbf{2.8} & \textbf{7.9} & \textbf{26.4} & \textbf{5.8} & \textbf{14.0} & \textbf{48.6} \\
Max-FLOPS      &    $2.8$ &    $7.9$ &   $26.1$ &    $7.2$ &   $17.9$ &   $49.1$ \\
Downclock      &    $2.9$ &    $8.5$ &   $27.6$ &   $12.0$ &   $25.3$ &   $57.8$ \\
Power-Capping  &    $3.1$ &   $18.7$ &  $465.4$ &   $11.8$ &   $33.9$ &   $77.6$ \\
Idle-GPUs      &   $44.8$ &   $71.5$ &   $89.3$ &  $330.6$ &  $369.9$ &  $386.3$ \\
\midrule
\multicolumn{7}{@{}l}{\textit{Conversation Workload}} \\[2pt]
\textbf{\slc{}} & \textbf{9.7} & \textbf{16.7} & \textbf{27.5} & \textbf{14.1} & \textbf{31.4} & \textbf{72.2} \\
Max-FLOPS      &    $9.6$ &   $17.0$ &   $30.0$ &   $18.4$ &   $66.2$ &  $126.9$ \\
Downclock      &   $10.4$ &   $19.5$ &   $38.4$ &   $27.4$ &   $78.4$ &  $137.0$ \\
Power-Capping  &   $11.6$ &   $36.9$ &  $184.6$ &   $87.6$ &  $153.3$ &  $260.4$ \\
Idle-GPUs      &  $156.8$ &  $217.6$ &  $276.0$ &  $474.1$ &  $539.4$ &  $597.4$ \\
\bottomrule
\end{tabular}
\end{table}

\raggedbottom
\parab{\slc{} robustness.} Each $2$-hour experiment at $175$~RPS generates $1.26$~M requests (P$99$ over $\sim${}$12$,$600$ samples); a percentile breakdown in Table\ref{tab:percentiles} confirms trends are consistent across P$95$/P$99$/P$99.9$. Forecast errors are mitigated by the $180$~s decision cycle, which provides five corrective opportunities per forecast interval; combined with sub-second \slc{} runtime, transient mismatches are absorbed within one cycle. Our power profile is deliberately challenging: a sustained ~$50\%$ drop at the largest site is rare in practice, where batteries, complementary renewables, or grid draw would buffer variability. All of our high-level findings are robust across multiple complementary power profiles that we have tested with.


\begin{figure}[t]
    \centering
    \includegraphics[width=0.8\columnwidth]{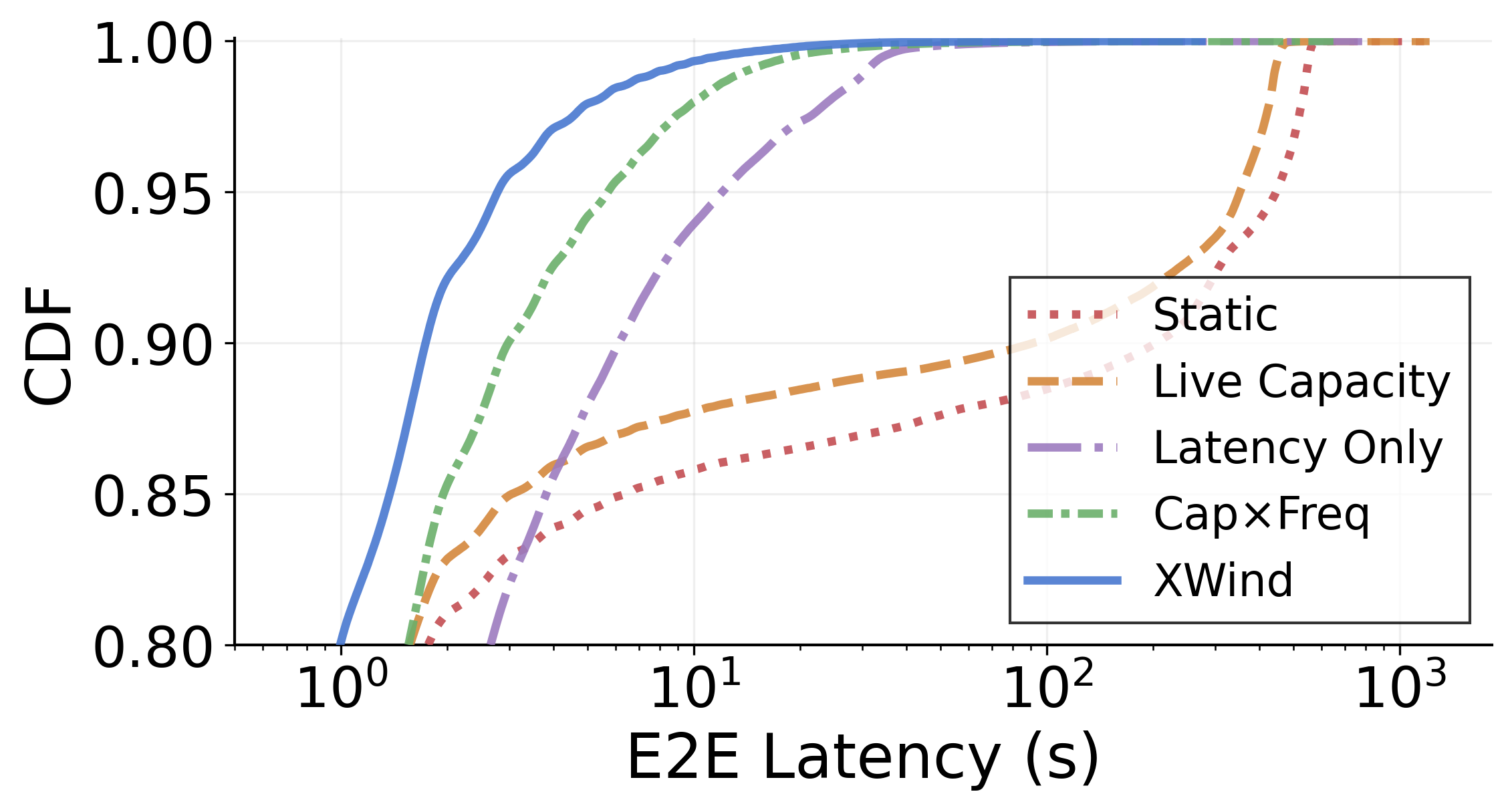}
    \caption{CDF of E$2$E latency for different routing strategies using the same reactive \slc{} (coding, $150$~RPS).}
    \label{fig:ablation}
\end{figure}

\subsection{\router{} Router Ablation Study}
\label{subsec:eval-ablation}
 
To isolate the contribution of each signal in \router{}'s cross-site routing (\textbf{Q$4$}), we evaluate four router variants, all paired with the same reactive \slc{} at each site, on the coding workload at $150$~RPS (Fig.~\ref{fig:ablation}). The variants differ \textit{only} in how the cross-site router computes routing weights.
 
\parab{Static routing is oblivious to power changes.} The \textit{Static} ($2$:$1$:$1$) router distributes traffic in proportion to site capacities, ignoring runtime signals entirely. Even when Site-$0$ loses $50\%$ of its power mid-trace, the router continues sending statically decided request volume there, causing significant queue buildup: P$99$ E$2$E: $546$~s, P$99$ queue-time: $540$~s.
 
\parab{Live capacity alone is insufficient.} \textit{Live Capacity} routing sets weights proportional to each site's active GPU count, adapting to \slc{}-driven reconfigurations. This yields only a $19\%$ P$99$ E$2$E improvement ($443$~s) over \textit{Static} routing: knowing \textit{how many} GPUs are active does not capture \textit{how congested} they are: a low-frequency site with a deep queue still attracts traffic proportional to its node count.
 
\parab{Latency feedback is the single largest lever.} \textit{Latency Only} routing weights sites inversely by EMA-smoothed TBT, with no capacity or frequency knowledge. This achieves $30.8$~s P$99$ E$2$E, $18\times$ lower than \textit{Static}, because latency directly reflects congestion: rising TBT under power contraction diverts traffic before queues build up. However, latency is a lagging indicator; by the time TBT increases, requests are already queued, thus inflating E$2$E latency.
 
\parab{Capacity$\times$frequency provides a leading signal.} \textit{Cap$\times$Freq}  router weighs each site by the product of active GPU count and operating frequency, capturing both quantity and quality of available compute. This yields P$99$ E$2$E of $14$~s ($2.2\times$ lower than \textit{Latency Only}) because the router \textit{anticipates} throughput changes at reconfiguration time rather than waiting for latency inflation to react.
 
\parab{\router{} combines leading and lagging signals.} The \router{} router integrates Cap$\times$Freq with EMA-smoothed per-site TBT latency, achieving $7.9$~s, a further $1.8\times$ improvement over \textit{Cap$\times$Freq} alone. Cap$\times$Freq signal sets the coarse routing weights based on each site's announced compute capacity, while the latency feedback loop applies fine-grained corrections that account for transient congestion not captured by the capacity signal. The CDF (Fig.~\ref{fig:ablation}) reveals that \textit{Cap$\times$Freq} and \router{} track closely below P$90$; the latency corrections primarily tighten the tail, where transient imbalances would otherwise cascade. Together, the two signals reduce P$99$ E2E by $69\times$ over \textit{Static} routing. 

\parab{}We ran similar experiments as above on a smaller H$100$ setup with $3$ sites ($4$, $2$, and $2$ GPUs). The high-level takeaways are same. The results have been omitted here for brevity.


\section{Related Work}

\parab{Carbon-aware and sustainable computing.} Previous works~\cite{carbonexplorer, ecovisor, pcrtm, carbonscaler, skybox, virtual_battery} shift workloads temporally or geographically to reduce carbon footprint but still rely on grid power. AI \greenf{} instead deploys compute directly at the renewable source, eliminating transmission losses, interconnection queues, and renewable energy credit abstractions while focusing on the AI inference workload.

\parab{Energy-efficient LLM serving.} DynamoLLM~\cite{dynamollm} minimizes AI inferencing energy by tuning parallelism and GPU frequencies, but assumes unconstrained power at a single site. VoltanaLLM~\cite{voltanallm} proposes frequency control and state-space routing for efficient LLM serving, also in a single-site grid-powered setting. 
Other GPU energy optimizations~\cite{zeus, ali2023perf_energy_gpu, dvfs_ai_accel} target training or accelerator-level efficiency using analytical models and fine-grained DVFS, not inference under power variability. \router{} instead dynamically routes the load between multiple sites based on fluctuating renewable power budgets.


\parab{Renewable-powered and modular data centers.} Prior work~\cite{parasol, greenpar, ren2012carbon, zhang2011greenware, govindan2011benefits} explored renewable-powered data centers using solar, batteries, and scheduling optimization. AI \greenf{} differs by: ($1$) targeting wind energy with its distinct variability; ($2$) right-sizing deployments at lower percentiles of peak generation to reduce power uncertainty; ($3$) exploiting power complementarity across dispersed sites; and ($4$) leveraging the stateless, request-level granularity of AI inferencing. Companies like Windcores~\cite{windcores}, Soluna~\cite{soluna}, and Westfalenwind~\cite{westfalenwind} deploy compute in wind farms for cryptocurrency and streaming; AI \greenf{} rather focuses on an emerging high-value AI workload.
\section{Discussions}

\parab{Logistics.} \greenf{} requires complex logistics: ROI/TCO analysis, footprint expansion planning, etc., all ongoing work with finance teams, renewable energy partners, and modular DC vendors. This paper rather focuses on the core technical contributions. Recent developments~\cite{nvidia_announcement, china_under_water} validate modular edge expansion; we target bringing these to renewable sites for AI inference.

\parab{Co-existence with traditional data centers.} \greenf{} could complement rather than replace conventional data centers. Requests are preferentially routed to \greenf{} sites for lower energy costs; overflow is absorbed by traditional DCs as elastic peak-load capacity. The router's per-site volume-to-latency mapping dynamically adjusts this allocation each cycle. Given the spatial complementarity and other guardrails, spillage is rare and needs minimal provisioning.

\parab{GPU phase-out.} With GPU leaders like NVIDIA announcing a new generation almost every year, 
hyperscalers need to come up with 
concrete plans to phase-out the quickly aging GPUs and make room for the newer ones to run training workloads. AI \greenf{} could also help address this mid-life GPU crisis, if needed, by shipping them to wind sites and running a fraction of the AI workload at lower CAPEX (already offset heavily at the full-fledged data center).

\parab{Elephants versus mice.} Asymmetric \greenf{} site sizes could cause routing imbalances under power variability. Our analysis (\S\ref{subsec:wind-capacity}) confirms that \azure{} regions host diverse wind farms within low network latency, allowing sizeable deployments that absorb power uncertainty; in rare cases, requests can spill over to the full-fledged data center.

\section{Conclusion}


This paper introduces AI \greenf{}, a deployment model that co-locates modular AI inference compute at wind farms to bypass grid delivery bottlenecks, and \router{}, a reactive cross-site router that serves LLM inference under variable renewable power without offline workload profiling. Evaluated on $64$ NVIDIA A$100$ GPUs across $3$ sites with \azure{} production traces, \router{} reduces P$99$ E$2$E latency by $22$–$52\%$ over the strongest dual-knob contender (Max-FLOPS) and by up to $98\%$ over single-knob baselines. 
\bibliographystyle{ACM-Reference-Format} 
\bibliography{main}
\end{document}